\documentclass[12pt]{article}
\usepackage{graphicx}
\usepackage{amsmath}
\usepackage{amssymb}
\usepackage{caption2}
\begin{document}
\bibliographystyle{prsty}
\begin{center}
{\large {\bf \sc{  Reanalysis of the mass spectrum of the scalar
hidden charm and hidden bottom tetraquark states }}} \\[2mm]
Zhi-Gang Wang \footnote{E-mail,wangzgyiti@yahoo.com.cn.  }     \\
 Department of Physics, North China Electric Power University,
Baoding 071003, P. R. China
\end{center}

\begin{abstract}
In this article, we study the mass spectrum of the scalar hidden
charm and hidden bottom tetraquark states which consist of the
axial-axial type and the vector-vector type diquark pairs with the
QCD sum rules.
\end{abstract}



\section{Introduction}

In 2007, a distinct peak ($Z(4430)$) was observed in the
$\pi^{\pm}\psi'$ invariant mass distribution near $4.43\,\rm{GeV}$
in the decays $B \rightarrow K \pi^{\pm} \psi'$  by the Belle
collaboration \cite{Belle-z4430}. The fitted Breit-Wigner   mass and
width are $M_Z=4433\pm 4  \pm 2 \,\rm{MeV}$ and $\Gamma_Z = 45
^{+18}_{-13} {}^{+30} _{-13} \,\rm{MeV}$. The statistical
significance of the observed peak is $6.5\,\sigma$. Using the same
data sample, the Belle collaboration also performed a full Dalitz
plot analysis  with a fitted  model that takes into account all the
known $K \pi$ resonances below $1780\,\rm{MeV}$ \cite{Belle0905}.
The significance of the fitted resonance is of $6.4\,\sigma$  and
agrees with the previous
 observation \cite{Belle-z4430}, the  updated parameters are $M_Z
=(4443^{+15}_{-12}{}^{+19}_{-13}) \,\rm{MeV}$ and  $\Gamma_Z =
(109^{+86}_{-43}{}^{+74}_{-56})\,\rm{MeV}$. However, the BaBar
collaboration do not confirm this resonance \cite{Babar0811}, i.e.
they observe no significant evidence for a $Z(4430)$ signal for any
of the processes investigated, neither in the total $J/\psi\pi$ or
$\psi'\pi$ mass distribution nor in the corresponding distributions
for the regions of $K\pi$ mass for which observation of the
$Z(4430)$ signal is reported. If the $Z(4430)$ exists  indeed, it
can't be a pure $c\bar{c}$ state due to the positive charge, and may
be an excellent   tetraquark state ($c\bar{c}u\bar{d}$) candidate
\cite{reviewXYZ,reviewXYZ2}.
 We can distinguish the multiquark states
 from the hybrids or charmonia with the criterion of
non-zero charge.

In 2008, the Belle collaboration  reported the first observation of
two resonance-like structures (the $Z(4050)$ and $Z(4250)$) in the
$\pi^+\chi_{c1}$ invariant mass distribution near $4.1 \,\rm{GeV}$
in the exclusive decays $\bar{B}^0\to K^- \pi^+ \chi_{c1}$
\cite{Belle-chipi}. Their quark contents must be some special
combinations of the $c\bar{c} u\bar{d}$, just like the $Z(4430)$,
they can't be the conventional mesons.  The $Z(4050)$ and $Z(4250)$
lie about $(0.5-0.6)\,\rm{GeV}$ above the $\pi^+\chi_{c1}$
threshold, the decay $ Z \to \pi^+\chi_{c1}$ can take place with the
"fall-apart" mechanism and it is OZI super-allowed, which can take
into account the large total width naturally. The spins of the
$Z(4050)$ and $Z(4250)$  are not determined yet, they can be scalar
or vector mesons. If they are scalar mesons, the decays $Z \to
\pi^+\chi_{c1}$ occur through the relative $P$-wave with the
phenomenological lagrangian $\mathcal {L}=g\chi^\alpha ( \pi
\partial_\alpha Z-Z\partial_\alpha \pi)$. On the other hand, if they
are vector mesons, the decays occur through the relative $S$-wave
with the phenomenological lagrangian $\mathcal {L}=g\chi^\alpha
Z_\alpha \pi $.  There have been several interpretations, such as
the tetraquark states \cite{ Wang0807,Wang08072,Ebert0808} and the
molecular states \cite{Xliu0808,Lee09,SLee08,GDing09}.

 In
Refs.\cite{Wang0807,Wang08072}, we assume  that the hidden charm
mesons $Z(4050)$ and $Z(4250)$ are vector (and scalar) tetraquark
states, and study their masses with the QCD sum rules. The numerical
results indicate that the mass of the vector hidden charm tetraquark
state is about $M_{Z}=(5.12\pm0.15)\,\rm{GeV}$ or
$M_{Z}=(5.16\pm0.16)\,\rm{GeV}$, and  the mass of the scalar hidden
charm tetraquark state
 is about $M_{Z}=(4.36\pm0.18)\,\rm{GeV}$. In
Refs.\cite{WangScalar,WangVector}, we study the mass spectrum of the
scalar and vector hidden charm and hidden bottom tetraquark states
using the QCD sum rules, and observe that the scalar hidden charm
tetraquark states may have smaller masses than the corresponding
vector states.  From our previous works, we can draw the conclusion
that the hidden charm meson $Z(4250)$ may be a scalar tetraquark
state \cite{Wang0807,Wang08072,WangScalar,WangVector}, although
other possibilities, such as a  hadro-charmonium resonance  and a
$D_1^+\bar{D}^0+ D^+\bar{D}_1^0$ molecular state  are not excluded.
We intend to study the mass spectrum of the scalar hidden charm and
hidden bottom tetraquark states which consist of diquark pairs
differ from our previous works.

The mass is a fundamental parameter in describing a hadron, whether
or not there exist those hidden charm and hidden bottom tetraquark
configurations is of great importance itself, because it provides a
new opportunity for a deeper understanding of the low energy QCD.

In Refs.\cite{Ebert0512,Ebert0812}, Ebert et al take the diquarks as
bound states of the light and heavy quarks in the color antitriplet
channel, and calculate their mass spectrum using a Schrodinger type
equation, then take the masses of the diquarks  as the basic input
parameters, and study the mass spectrum of the heavy tetraquark
states as bound states of the diquark-antidiquark system. In
Refs.\cite{Maiani20042,Maiani2008,Polosa0902}, Maiani et al take the
diquarks as the basic constituents, examine the rich spectrum of the
diquark-antidiquark states  with  the constituent diquark masses and
the spin-spin
 interactions, and try to  accommodate some of the newly observed charmonium-like resonances not
 fitting a pure $c\bar{c}$ assignment. In Ref.\cite{Zouzou86}, Zouzou et al  solve the four-body ($\bar{Q}
\bar{Q}qq$) problem by three different variational methods with a
non-relativistic potential considering explicitly virtual
meson-meson components in the wave-functions, search for possible
bound states below the threshold for the spontaneous dissociation
into two mesons, and observe that the exotic bound states
$\bar{Q}\bar{Q}qq$ maybe exist for  unequal quark masses (the ratio
$m_Q/m_q$ is large enough). The studies  using a potential derived
from the MIT bag model in the Born-Oppenheimer approximation support
this observation \cite{Heller87,Heller88}. In Ref.\cite{Manohar93},
Manohar and Wise study systems of two heavy-light  mesons
interacting through  an one-pion exchange potential determined by
the heavy meson chiral perturbation theory and observe the long
range potential maybe sufficiently attractive to produce a weakly
bound two-meson state in the case $Q=b$. In
Ref.\cite{Silvestre-Brac}, the $L=0$ tetraquark states
$QQ\overline{QQ}$ ($Q$ denotes both $Q$ and $q$) are analyzed in  a
chromo-magnetic model where only a constant hyperfine potential   is
retained.

In this article, we re-study the mass spectrum of the scalar hidden
charm and hidden bottom tetraquark states using  the QCD sum rules
\cite{SVZ79,Reinders85}.  In the QCD sum rules, the operator product
expansion is used to expand the time-ordered currents into a series
of quark and gluon condensates which parameterize the long distance
properties of  the QCD vacuum. Based on the quark-hadron duality, we
can obtain copious information about the hadronic parameters at the
phenomenological side \cite{SVZ79,Reinders85}.

 The hidden charm and hidden bottom tetraquark states ($Z$) have the symbolic
quark structures:
\begin{align}
  Z^+ = Q\bar{Q} u  \bar{d}  ;~~~~
  Z^0 = \frac{1}{\sqrt{2}}Q\bar{Q}&( u  \bar{u}-d  \bar{d})  ;~~~~
  Z^- =Q\bar{Q}d\bar{u}    ; \nonumber\\
  Z_s^+ = Q\bar{Q}u  \bar{s} ;~~~~
  Z_s^- =Q\bar{Q}  s\bar{u}  ;&~~~~
  Z_s^0 = Q\bar{Q}d  \bar{s} ;~~~~
  \overline Z_s^0 = Q\bar{Q}s\bar{d} ; \nonumber \\
  Z_\varphi= \frac{1}{\sqrt{2}} Q\bar{Q} (u\bar{u}+d\bar{d});
  &~~~~Z_\phi =  Q\bar{Q} s\bar{ s} \, ,
\end{align}
where the $Q$ denotes  the heavy quarks $c$ and $b$.

We take the diquarks as the basic constituents to study the
tetraquark states
 following Jaffe and Wilczek \cite{Jaffe2003,Jaffe2004}.
 The heavy tetraquark system could be described  by a double-well
potential with  two light quarks $q'\bar{q}$ lying in the two wells
respectively.  In the heavy quark limit, the $c$ (and $b$) quark can
be taken  as a static well potential, which binds the light quark
$q$ to form a diquark in the color antitriplet channel.  The
attractive interactions of one-gluon exchange  favor  formation of
the diquarks in  color antitriplet $\overline{3}_{ c}$, flavor
antitriplet $\overline{3}_{ f}$ and spin singlet $1_s$
\cite{GI1,GI2}. The diquarks have five Dirac tensor structures,
scalar $C\gamma_5$, pseudoscalar $C$, vector $C\gamma_\mu \gamma_5$,
axial vector $C\gamma_\mu $  and tensor $C\sigma_{\mu\nu}$. The
structures $C\gamma_\mu $ and $C\sigma_{\mu\nu}$ are symmetric, the
structures $C\gamma_5$, $C$ and $C\gamma_\mu \gamma_5$ are
antisymmetric.    In Refs.\cite{Wang08072,WangScalar}, we assume the
scalar hidden charm and hidden bottom mesons $Z$ consist of  the
$C\gamma_5-C \gamma_5$ type diquark structures rather than the $C-C
$ type diquark structures, and observe that the $C\gamma_5-C
\gamma_5$ type tetraquark states have much smaller masses than the
corresponding $C-C $ type tetraquark states; our numerical results
of the $C-C$ type tetraquark states will be presented elsewhere.

In this article, we assume the scalar hidden charm and hidden bottom
tetraquak states which consist of the $C\gamma_\mu -C\gamma^\mu $
type and the $C\gamma_\mu \gamma_5-C\gamma^\mu \gamma_5$ type
diqaurk pairs and study the mass spectrum. Naively, we expect the
$C\gamma_\mu -C\gamma^\mu $ type and the $C\gamma_\mu
\gamma_5-C\gamma^\mu \gamma_5$ type tetraquark states have larger
masses than  the corresponding  $C\gamma_5 -C\gamma_5 $ type
tetraquark states.

The article is arranged as follows:  we derive the QCD sum rules for
  the scalar hidden charm and hidden bottom tetraquark states $Z$  in section 2; in section 3, numerical
results and discussions; section 4 is reserved for conclusion.

\section{QCD sum rules for  the scalar tetraquark states $Z$ }
In the following, we write down  the two-point correlation functions
$\Pi(p)$  in the QCD sum rules,
\begin{eqnarray}
\Pi(p)&=&i\int d^4x e^{ip \cdot x} \langle
0|T\left\{J/\eta(x)J/\eta^{\dagger}(0)\right\}|0\rangle \, ,
\end{eqnarray}
where the  $J(x)$ and $\eta(x)$ denotes the interpolating currents
$J_{Z^+}(x)$, $J_{Z^0}(x)$,  $\cdots$, $\eta_{Z^+}(x)$,
$\eta_{Z^0}(x)$, $\cdots$,
\begin{eqnarray}
J_{Z^+}(x)&=& \epsilon^{ijk}\epsilon^{imn}u_j^T(x) C\gamma_\mu
Q_k(x) \bar{Q}_m(x) \gamma^\mu  C \bar{d}_n^T(x)\, , \nonumber\\
J_{Z^0}(x)&=& \frac{\epsilon^{ijk}\epsilon^{imn}}{\sqrt{2}}
\left[u_j^T(x) C\gamma_\mu Q_k(x) \bar{Q}_m(x) \gamma^\mu  C
\bar{u}_n^T(x)-(u\rightarrow d)\right]\, , \nonumber\\
J_{Z^+_s}(x)&=& \epsilon^{ijk}\epsilon^{imn}u_j^T(x) C\gamma_\mu
Q_k(x)\bar{Q}_m(x) \gamma^\mu  C \bar{s}_n^T(x)\, , \nonumber\\
J_{Z^0_s}(x)&=& \epsilon^{ijk}\epsilon^{imn}d_j^T(x) C\gamma_\mu
Q_k(x)\bar{Q}_m(x) \gamma^\mu  C \bar{s}_n^T(x)\, , \nonumber\\
J_{Z_\varphi}(x)&=& \frac{\epsilon^{ijk}\epsilon^{imn}}{\sqrt{2}}
\left[u_j^T(x) C\gamma_\mu Q_k(x) \bar{Q}_m(x) \gamma^\mu  C
\bar{u}_n^T(x) +(u\rightarrow d)\right]\, , \nonumber\\
 J_{Z_\phi}(x)&=& \epsilon^{ijk}\epsilon^{imn}s_j^T(x) C\gamma_\mu
Q_k(x)\bar{Q}_m(x) \gamma^\mu  C \bar{s}_n^T(x)\, , \nonumber \\
\eta_{Z^+}(x)&=& \epsilon^{ijk}\epsilon^{imn}u_j^T(x)
C\gamma_\mu\gamma_5
Q_k(x) \bar{Q}_m(x) \gamma_5\gamma^\mu  C \bar{d}_n^T(x)\, , \nonumber\\
\eta_{Z^0}(x)&=& \frac{\epsilon^{ijk}\epsilon^{imn}}{\sqrt{2}}
\left[u_j^T(x) C\gamma_\mu \gamma_5Q_k(x) \bar{Q}_m(x)
\gamma_5\gamma^\mu  C
\bar{u}_n^T(x)-(u\rightarrow d)\right]\, , \nonumber\\
\eta_{Z^+_s}(x)&=& \epsilon^{ijk}\epsilon^{imn}u_j^T(x)
C\gamma_\mu\gamma_5
Q_k(x)\bar{Q}_m(x) \gamma_5\gamma^\mu  C \bar{s}_n^T(x)\, , \nonumber\\
\eta_{Z^0_s}(x)&=& \epsilon^{ijk}\epsilon^{imn}d_j^T(x)
C\gamma_\mu\gamma_5
Q_k(x)\bar{Q}_m(x) \gamma_5\gamma^\mu  C \bar{s}_n^T(x)\, , \nonumber\\
\eta_{Z_\varphi}(x)&=& \frac{\epsilon^{ijk}\epsilon^{imn}}{\sqrt{2}}
\left[u_j^T(x) C\gamma_\mu \gamma_5Q_k(x) \bar{Q}_m(x)
\gamma_5\gamma^\mu  C
\bar{u}_n^T(x) +(u\rightarrow d)\right]\, , \nonumber\\
 \eta_{Z_\phi}(x)&=& \epsilon^{ijk}\epsilon^{imn}s_j^T(x) C\gamma_\mu\gamma_5
Q_k(x)\bar{Q}_m(x)\gamma_5 \gamma^\mu  C \bar{s}_n^T(x)\, ,
\end{eqnarray}
where the $i$, $j$, $k$, $\cdots$  are color indexes. In the isospin
limit, the interpolating currents result in six distinct expressions
for the correlation functions $\Pi(p)$ , which are characterized by
the number of the $s$ quark they contain. In
Refs.\cite{WangScalar,WangVector}, we observe that the ground state
masses of the scalar and vector tetraquarks are characterized  by
the number of the $s$ quarks they contain, $M_{0}\leq M_{s}\leq
M_{ss}$; the energy gap between $M_{0}$ and $M_{ss}$ is about
$(0.05-0.15)\,\rm{GeV}$. In this article, we study the interpolating
currents which contains zero and two $s$ quarks for simplicity.

We can insert  a complete set of intermediate hadronic states with
the same quantum numbers as the current operators $J(x)$ and
$\eta(x)$ into the correlation functions  $\Pi(p)$  to obtain the
hadronic representation \cite{SVZ79,Reinders85}. After isolating the
ground state contribution from the pole terms  of the $Z$, we get
the following result,
\begin{eqnarray}
\Pi(p)&=&\frac{\lambda_{Z}^2}{M_{Z}^2-p^2} +\cdots \, \, ,
\end{eqnarray}
where the pole residue (or coupling) $\lambda_Z$ is defined by
\begin{eqnarray}
\lambda_{Z}  &=& \langle 0|J/\eta(0)|Z(p)\rangle \, .
\end{eqnarray}

The  contributions from the two-particle and many-particle reducible
states are supposed to be small enough to be neglected safely, for
example, the scattering state $\chi_{c1}\pi^+$ in the
$\bar{c}c\bar{d}u$ channel,
\begin{eqnarray}
\Pi(p)&=&i\lambda_{\chi_{c1}\pi^+}^2 \int \frac{d^4q}{(2\pi)^4}
\frac{p^\mu p^\nu
}{\left[q^2-m_{\chi_{c1}}^2\right]\left[(p-q)^2-m_{\pi}^2\right]}\left[-g_{\mu\nu}+\frac{q_\mu
q_\nu}{m_{\chi_{c1}}^2} \right]+\cdots \, ,
\end{eqnarray}
where
\begin{eqnarray}
\langle 0|J_{Z^+}(0)|\chi_{c1}\pi^+\rangle &=&
\lambda_{\chi_{c1}\pi^+} p_\mu \epsilon^\mu \, ,
\end{eqnarray}
the $\epsilon_\mu$ is the polarization vector of the axial-vector
meson $\chi_{c1}$. We can estimate the coupling
$\lambda_{\chi_{c1}\pi^+}$ with the soft $\pi$ theorem,
\begin{eqnarray}
\langle 0|J_{Z^+}(0)|\chi_{c1}\pi^+\rangle
&=&-\frac{i}{f_\pi}\langle 0|\left[Q_5,J_{Z^+}(0)\right]|\chi_{c1}
\rangle \, , \nonumber \\
&=&\frac{i}{f_\pi}\langle 0|J_{P}(0)|\chi_{c1}\rangle
=-\frac{\lambda_Pp_\mu \epsilon^\mu}{f_\pi}\, ,
\end{eqnarray}
where
\begin{eqnarray}
Q_5&=&\int d^3x u^+(x)i\gamma_5d(x) \, , \nonumber\\
J_P(x)&=&\epsilon^{ijk}\epsilon^{imn} \left[u_j^T(x) C\gamma_\mu
\gamma_5 c_k(x) \bar{c}_m(x) \gamma^\mu  C \bar{u}_n^T(x) \right.
\nonumber \\
&& \left. +d_j^T(x) C\gamma_\mu  c_k(x) \bar{c}_m(x)
\gamma_5\gamma^\mu  C \bar{d}_n^T(x) \right] \, .
\end{eqnarray}
As the main Fock states of the charmonia are the $\bar{c}c$
components, the coupling $\lambda_P$ between the pseudoscalar
tetraquark current $J_P(x)$ and the axial-vector meson $\chi_{c1}$
should be very small.

We can perform  Fierz re-ordering in both the Dirac spin space and
the color space to express the tetraquark current $J_{Z^+}(x)$ in
the following form,
\begin{eqnarray}
J_{Z^+}(x)&=&\frac{3}{8}\bar{d}(x)u(x)\bar{c}(x)c(x)+\frac{3}{8}\bar{d}(x)i\gamma_5u(x)\bar{c}(x)i\gamma_5c(x)
+\frac{3}{16}\bar{d}(x)\gamma_\alpha u(x)\bar{c}(x)\gamma^\alpha
c(x) \nonumber \\
&&-\frac{3}{16}\bar{d}(x)\gamma_\alpha\gamma_5
u(x)\bar{c}(x)\gamma^\alpha \gamma_5
c(x)-\frac{1}{2}\bar{d}(x)\frac{\lambda^i}{2}u(x)\bar{c}(x)\frac{\lambda^i}{2}c(x)\nonumber\\
&&-\frac{1}{2}\bar{d}(x)i\gamma_5\frac{\lambda^i}{2}u(x)\bar{c}(x)i\gamma_5\frac{\lambda^i}{2}c(x)
-\frac{1}{4}\bar{d}(x)\gamma_\alpha
\frac{\lambda^i}{2}u(x)\bar{c}(x)\gamma^\alpha\frac{\lambda^i}{2}
c(x)\nonumber \\
&&+\frac{1}{4}\bar{d}(x)\gamma_\alpha\gamma_5
\frac{\lambda^i}{2}u(x)\bar{c}(x)\gamma^\alpha \gamma_5
\frac{\lambda^i}{2}c(x) \, ,
\end{eqnarray}
where the $\lambda^i$ are the matrix elements of the $SU(2)$ group
in adjoint representation, the $i=1,2,3$ are the color indexes. The
scalar tetraquark current which consists  of a axial-vector diquark
pair is a special composition of the $S-S$, $P-P$, $V-V$, $A-A$,
$S^i-S^i$, $P^i-P^i$, $V^i-V^i$ and $A^i-A^i$ color-singlet and
color-triplet meson-meson  type currents, the $S$, $P$, $V$ and $A$
denote the scalar, pseudoscalar, vector and axial-vector
respectively. The color-singlet meson-meson type currents, for
example, $\bar{d}(x)\gamma_\alpha\gamma_5
u(x)\bar{c}(x)\gamma^\alpha \gamma_5 c(x)$,
$\bar{d}(x)i\gamma_5u(x)\bar{c}(x)i\gamma_5c(x)$, have very small
two-particle reducible contributions \cite{Lee2005}.

 After performing the standard procedure of the QCD sum rules, we obtain the following  four   sum rules for the
 interpolating currents contain two $s$  quarks:
\begin{eqnarray}
\lambda_{Z}^2 e^{-\frac{M_Z^2}{M^2}}= \int_{\Delta}^{s_0} ds
\rho_{\pm}(s)e^{-\frac{s}{M^2}} \, ,
\end{eqnarray}
the explicit expressions of the spectral densities $\rho_{\pm}(s)$
are  presented in  the appendix, the $+$ and $-$ denote the
$C\gamma_\mu-C\gamma^\mu$ type and the
$C\gamma_\mu\gamma_5-C\gamma^\mu\gamma_5$ type interpolating
currents respectively;  the $s_0$ is the continuum threshold
parameter and the $M^2$ is the Borel  parameter.
       We can obtain  four  sum rules in  the
$c\bar{c}q\bar{q}$ and $b\bar{b}q\bar{q}$ channels with a simple
replacement $m_s\rightarrow m_q$,  $\langle
\bar{s}s\rangle\rightarrow\langle \bar{q}q\rangle$ and $\langle
\bar{s}g_s \sigma Gs\rangle\rightarrow\langle \bar{q}g_s \sigma
Gq\rangle$.

 We carry out the operator
product expansion to the vacuum condensates adding up to
dimension-10. In calculation, we
 take  assumption of vacuum saturation for  high
dimension vacuum condensates, they  are always
 factorized to lower condensates with vacuum saturation in the QCD sum rules,
  factorization works well in  large $N_c$ limit.
In this article, we take into account the contributions from the
quark condensates,  mixed condensates, and neglect the contributions
from the gluon condensate. The contributions  from the gluon
condensates  are suppressed by large denominators and would not play
any significant roles for the light tetraquark states
\cite{Wang1,Wang2}, the heavy tetraquark state \cite{Wang08072} and
the  heavy molecular state \cite{Wang0904}.  There are many terms
involving the gluon condensate for the heavy tetraquark states and
heavy molecular states in the operator product expansion (one can
consult Refs.\cite{Wang08072,Wang0904,WangZhang} for example), we
neglect the gluon condensates for simplicity.

 Differentiate  the Eq.(11) with respect to  $\frac{1}{M^2}$, then eliminate the
 pole residues $\lambda_{Z}$, we can obtain the  sum rules  for
 the masses  of the tetraquark quark  states $Z$,
 \begin{eqnarray}
 M_Z^2= \frac{\int_{\Delta}^{s_0} ds
\frac{d}{d(-1/M^2)}\rho_{\pm}(s)e^{-\frac{s}{M^2}}
}{\int_{\Delta}^{s_0} ds \rho_{\pm}(s)e^{-\frac{s}{M^2}}}\, .
\end{eqnarray}

\section{Numerical results and discussions}
The input parameters are taken to be the standard values $\langle
\bar{q}q \rangle=-(0.24\pm 0.01 \,\rm{GeV})^3$, $\langle \bar{s}s
\rangle=(0.8\pm 0.2 )\langle \bar{q}q \rangle$, $\langle
\bar{q}g_s\sigma Gq \rangle=m_0^2\langle \bar{q}q \rangle$, $\langle
\bar{s}g_s\sigma Gs \rangle=m_0^2\langle \bar{s}s \rangle$,
$m_0^2=(0.8 \pm 0.2)\,\rm{GeV}^2$,  $m_s=(0.14\pm0.01)\,\rm{GeV}$,
$m_u=m_d\approx0$, $m_c=(1.35\pm0.10)\,\rm{GeV}$ and
$m_b=(4.8\pm0.1)\,\rm{GeV}$ at the energy scale  $\mu=1\, \rm{GeV}$
\cite{SVZ79,Reinders85,Ioffe2005}.

In the conventional QCD sum rules \cite{SVZ79,Reinders85}, there are
two criteria (pole dominance and convergence of the operator product
expansion) for choosing  the Borel parameter $M^2$ and threshold
parameter $s_0$. We impose the two criteria on the heavy tetraquark
states to choose the Borel parameter $M^2$ and threshold parameter
$s_0$.

In Refs.\cite{Wang08072,WangScalar}, we assume that  the
resonance-like structures $Z(4050)$ and $Z(4250)$    are  scalar
tetraquark states which consist of the scalar diquark pairs, and
take the threshold parameter tentatively as $s_0=(4.248+0.5)^2\,
\rm{GeV}^2\approx 23 \, \rm{GeV}^2$ to take into account all
possible contributions from the ground states,  where  the energy
gap between the ground states and the first radial excited states is
chosen  to  be $0.5\,\rm{GeV}$. Then we take into account the
$SU(3)$ symmetry of the light flavor quarks and the mass difference
between the heavy quarks,  choose other  threshold parameters
tentatively, and use those values as a guide to determine the
threshold parameters $s_0$ with the QCD sum rules.

In this article, we study the scalar hidden charm and hidden bottom
tetraquark states which consist of the axial-axial type and the
vector-vector type diquark pairs, and search for other possible
tetraquark structures of the resonance-like states $Z(4050)$ and
$Z(4250)$. Naively, we expect the $C\gamma_\mu \gamma_5-C\gamma^\mu
\gamma_5$ type and the $C\gamma_\mu -C\gamma^\mu $ type tetraquark
states have larger masses than the corresponding   $C\gamma_5
-C\gamma_5 $ type tetraquark states, and use the threshold
parameters  in Ref.\cite{WangScalar} as a  guide  to determine the
threshold parameters $s_0$ with the QCD sum rules.

The contributions from the high dimension vacuum condensates  in the
operator product expansion are shown in Figs.1-2, where (and
thereafter) we  use the $\langle\bar{q}q\rangle$ to denote the quark
condensates $\langle\bar{q}q\rangle$, $\langle\bar{s}s\rangle$ and
the $\langle\bar{q}g_s \sigma Gq\rangle$ to denote the mixed
condensates $\langle\bar{q}g_s \sigma Gq\rangle$, $\langle\bar{s}g_s
\sigma Gs\rangle$. From the figures, we can see that  the
contributions from the high dimension condensates change quickly
with variation of the Borel parameter at the values $M^2\leq 2.6
\,\rm{GeV}^2\,(2.8\,\rm{GeV}^2)$ and $M^2\leq 7.2
\,\rm{GeV}^2\,(7.6\,\rm{GeV}^2)$ in the hidden charm and hidden
bottom channels respectively for the $C\gamma_\mu-C\gamma^\mu$
($C\gamma_\mu\gamma_5-C\gamma^\mu\gamma_5$) type interpolating
currents, such an unstable behavior cannot lead to stable sum rules,
our numerical results confirm this conjecture, see Fig.4.

At the values $M^2\geq 2.6\,\rm{GeV}^2 \,(2.8\,\rm{GeV}^2)$ and
$s_0\geq 23\,\rm{GeV}^2\,(27\,\rm{GeV}^2)$, the contributions from
the $\langle \bar{q}q\rangle^2+\langle \bar{q}q\rangle \langle
\bar{q}g_s \sigma Gq\rangle $ term are less than  $14\%\,(23.5\%)$
in the $c\bar{c}q\bar{q}$ channel, the corresponding contributions
are less than  $4\%\,(13\%)$ in the $c\bar{c}s\bar{s}$ channels; the
contributions from the vacuum condensate of the highest dimension
$\langle\bar{q}g_s \sigma Gq\rangle^2$ are less than
$2.5\%\,(2.5\%)$ and $1.5\%\,(3\%)$ in the $c\bar{c}q\bar{q}$ and
$c\bar{c}s\bar{s}$ channels respectively; we expect the operator
product expansion is convergent  for the $C\gamma_\mu -C\gamma^\mu $
($C\gamma_\mu\gamma_5 -C\gamma^\mu\gamma_5 $) type interpolating
currents in the hidden charm channels,.

At the values $M^2\geq 7.2\,\rm{GeV}^2 \,(7.6\,\rm{GeV}^2)$ and
$s_0\geq 136\,\rm{GeV}^2\,(146\,\rm{GeV}^2)$, the contributions from
the $\langle \bar{q}q\rangle^2+\langle \bar{q}q\rangle \langle
\bar{q}g_s \sigma Gq\rangle $ term are less than  $10.5\%\,(19\%)$
in the $b\bar{b}q\bar{q}$ channel, the corresponding contributions
are less than  $3.5\%\,(8.5\%)$ in the $b\bar{b}s\bar{s}$ channels;
the contributions from the vacuum condensate of the highest
dimension $\langle\bar{q}g_s \sigma Gq\rangle^2$ are less than
$5\%\,(6\%)$ and $3\%\,(6\%)$ in  the $b\bar{b}q\bar{q}$ and
$b\bar{b}s\bar{s}$ channels respectively; we also expect the
operator product expansion is convergent  for the $C\gamma_\mu
-C\gamma^\mu $ ($C\gamma_\mu\gamma_5 -C\gamma^\mu\gamma_5 $) type
interpolating currents in the hidden bottom channels.

 In this article, we take the uniform Borel parameter
$M^2_{min}$, i.e. $M^2_{min}\geq 2.6 \, \rm{GeV}^2$ $(2.8 \,
\rm{GeV}^2)$ and $M^2_{min}\geq 7.2 \, \rm{GeV}^2\,(7.6 \,
\rm{GeV}^2)$ in the hidden charm  and hidden bottom channels
respectively for the $C\gamma_\mu -C\gamma^\mu $
($C\gamma_\mu\gamma_5 -C\gamma^\mu\gamma_5 $) type interpolating
currents.

In Fig.3, we show the  contributions from the pole terms with
variation of the Borel parameters and the threshold parameters. The
pole contributions are larger than (or equal) $50\%\,(52\%)$ at the
value $M^2 \leq 3.2 \, \rm{GeV}^2 $ and $s_0\geq
23\,\rm{GeV}^2\,(27\,\rm{GeV}^2),$ $24\,\rm{GeV}^2 $
$(28\,\rm{GeV}^2)$ in the $c\bar{c}q\bar{q}$,
   $c\bar{c}s\bar{s}$
channels respectively, and larger than (or equal) $51\%\,(52\%)$ at
the value $M^2 \leq 8.2 \, \rm{GeV}^2 $ and $s_0\geq
136\,\rm{GeV}^2\,(146\,\rm{GeV}^2),$ $138\,\rm{GeV}^2 $
$(148\,\rm{GeV}^2)$ in  the $b\bar{b}q\bar{q}$ and
$b\bar{b}s\bar{s}$ channels respectively for the the $C\gamma_\mu
-C\gamma^\mu $ ($C\gamma_\mu\gamma_5 -C\gamma^\mu\gamma_5 $) type
interpolating currents. Again we take the uniform Borel parameter
$M^2_{max}$, i.e. $M^2_{max}\leq 3.2 \, \rm{GeV}^2$ and
$M^2_{max}\leq 8.2 \, \rm{GeV}^2$ in the hidden charm  and hidden
bottom channels respectively.

For the $C\gamma_\mu-C\gamma^\mu$ type interpolating currents, the
threshold parameters are taken as $s_0=(24\pm1)\,\rm{GeV}^2$,
$(25\pm1)\,\rm{GeV}^2$, $(138\pm2)\,\rm{GeV}^2$,  and
$(140\pm2)\,\rm{GeV}^2$ in the $c\bar{c}q\bar{q}$,
    $c\bar{c}s\bar{s}$, $b\bar{b}q\bar{q}$,
    and $b\bar{b}s\bar{s}$ channels respectively;
   the Borel parameters are taken as $M^2=(2.6-3.2)\,\rm{GeV}^2$ and
   $(7.2-8.2)\,\rm{GeV}^2$ in the
hidden charm and hidden bottom channels respectively.

For the $C\gamma_\mu\gamma_5-C\gamma^\mu\gamma_5$ type interpolating
currents, the threshold parameters are taken as
$s_0=(28\pm1)\,\rm{GeV}^2$, $(29\pm1)\,\rm{GeV}^2$,
$(148\pm2)\,\rm{GeV}^2$,  and $(150\pm2)\,\rm{GeV}^2$ for the
$c\bar{c}q\bar{q}$,
    $c\bar{c}s\bar{s}$, $b\bar{b}q\bar{q}$,
    and $b\bar{b}s\bar{s}$ channels, respectively;
   the Borel parameters are taken as $M^2=(2.8-3.2)\,\rm{GeV}^2$ and
   $(7.6-8.2)\,\rm{GeV}^2$ in the
hidden charm and hidden bottom channels respectively.

      In those regions,  the pole contributions are about  $(47-75)\%$,
$(51-78)\%$, $(51-70)\%$ and $(53-72)\%$ in  the $c\bar{c}q\bar{q}$,
    $c\bar{c}s\bar{s}$, $b\bar{b}q\bar{q}$
     and $b\bar{b}s\bar{s}$ channels respectively for the
$C\gamma_\mu-C\gamma^\mu$ type interpolating currents; while
      the pole contributions are about  $(52-75)\%$,
$(52-74)\%$, $(52-68)\%$ and $(52-67)\%$ in  the $c\bar{c}q\bar{q}$,
    $c\bar{c}s\bar{s}$, $b\bar{b}q\bar{q}$
     and $b\bar{b}s\bar{s}$ channels respectively for $C\gamma_\mu\gamma_5-C\gamma^\mu\gamma_5$ type
     interpolating currents;   the two criteria of the QCD sum rules
are fully  satisfied  \cite{SVZ79,Reinders85}.

If we take  uniform pole contributions,   the interpolating current
with more $s$ quarks requires slightly larger threshold parameter
due to the $SU(3)$ breaking effects, see Fig.3. The threshold
parameters in the $c\bar{c}q\bar{q}$ and $b\bar{b}q\bar{q}$ channels
are slightly smaller than the corresponding ones in the
$c\bar{c}s\bar{s}$ and $b\bar{b}s\bar{s}$ channels respectively.
Naively, we expect the tetraquark  state with more $s$ quarks will
have larger mass, our numerical calculations confirm this
conjecture, see Fig.4. In that figure we plot the  tetraquark  state
masses $M_Z$ with variation of the Borel parameters and the
threshold parameters.

The Borel windows $M_{max}^2-M_{min}^2$ change with  variations of
the  threshold parameters $s_0$, see Fig.3. In this article, the
Borel windows  are taken as  $0.6\,\rm{GeV}^2\,(0.4\,\rm{GeV}^2)$
and $1.0\,\rm{GeV}^2\,(0.6\,\rm{GeV}^2)$ in the hidden charm and
hidden bottom channels respectively for the
$C\gamma_\mu-C\gamma^\mu$
($C\gamma_\mu\gamma_5-C\gamma^\mu\gamma_5$) type
     interpolating currents; they are small enough.
Furthermore, we take uniform Borel windows and smear the dependence
on the threshold parameters $s_0$ in each channel.  If we take
larger threshold parameters,  the Borel windows are larger and the
resulting  masses are larger, see Fig.4. In this article, we intend
to  calculate the possibly  lowest masses which are supposed to be
the ground state masses  by imposing the two criteria of the QCD sum
rules.

Taking into account all uncertainties of the input parameters,
finally we obtain the values of the masses and pole resides of
 the scalar tetraquark states  $Z$, which are  shown in Figs.5-6 and Tables 1-2.
In this article,  we calculate the uncertainties $\delta$  with the
formula
\begin{eqnarray}
\delta=\sqrt{\sum_i\left(\frac{\partial f}{\partial
x_i}\right)^2\mid_{x_i=\bar{x}_i} (x_i-\bar{x}_i)^2}\,  ,
\end{eqnarray}
 where the $f$ denote  the
hadron mass  $M_Z$ and the pole residue $\lambda_Z$,  the $x_i$
denote the input QCD parameters $m_c$, $m_b$, $\langle \bar{q}q
\rangle$, $\langle \bar{s}s \rangle$, $\cdots$. As the partial
 derivatives   $\frac{\partial f}{\partial x_i}$ are difficult to carry
out analytically, we take the  approximation $\left(\frac{\partial
f}{\partial x_i}\right)^2 (x_i-\bar{x}_i)^2\approx
\left[f(\bar{x}_i\pm \Delta x_i)-f(\bar{x}_i)\right]^2$ in the
numerical calculations.

From Tables 1-2, we can see that the uncertainties of the masses
$M_Z$ are rather small (about $(3-4)\%$ in the hidden charm channels
and $(1-2)\%$ in the hidden bottom channels),  while the
uncertainties of the pole residues $\lambda_{Z}$ are rather large
(about $(20-40)\%$). The uncertainties of the input parameters
($\langle \bar{q}q \rangle$, $\langle \bar{s}s \rangle$, $\langle
\bar{s}g_s\sigma G s \rangle$, $\langle \bar{q}g_s\sigma G q
\rangle$,  $m_s$,  $m_c$ and $m_b$) vary in the range $(2-25)\%$,
the uncertainties of the pole  residues $\lambda_{Z}$ are
reasonable. We obtain the  squared masses   $M_Z^2$ through a
fraction, see Eq.(12), the uncertainties in the numerator and
denominator which origin from a given input parameter (for example,
$\langle \bar{s}s \rangle$, $\langle \bar{s}g_s\sigma G s \rangle$)
cancel out with each other, and result in small net uncertainty.

\begin{table}
\begin{center}
\begin{tabular}{|c|c|c|c|c|c|}
\hline\hline tetraquark states &$C\gamma_\mu\gamma_5-C\gamma^\mu\gamma_5$&$C\gamma_\mu-C\gamma^\mu$& $C\gamma_5-C\gamma_5$ & Refs.\cite{Ebert0512,Ebert0812}\\
\hline
      $c\bar{c}s\bar{s}$  &$4.82\pm0.14$ &$4.45\pm0.16$&$4.44\pm0.16$ &$4.110$\\ \hline
            $c\bar{c}q\bar{q} $& $4.56\pm0.14$& $4.36\pm0.18$&$4.37\pm0.18$& $3.852$\\      \hline
    $b\bar{b}s\bar{s}$  & $11.70\pm0.18$&$11.23\pm0.16$&$11.31\pm0.16$ &$11.133$\\ \hline
            $ b\bar{b}q\bar{q} $&$11.38\pm0.13$ &$11.14\pm0.19$ &$11.27\pm0.20$ &$10.942$ \\      \hline
    \hline
\end{tabular}
\end{center}
\caption{ The masses (in unit of GeV)  of the scalar tetraquark
states, the values for the $C\gamma_5-C\gamma_5$ type scalar
tetraquark states are taken from Ref.\cite{WangScalar}. }
\end{table}

\begin{table}
\begin{center}
\begin{tabular}{|c|c|c|}
\hline\hline tetraquark states&$C\gamma_\mu\gamma_5-C\gamma^\mu\gamma_5$ &$C\gamma_\mu-C\gamma^\mu$\\
\hline
      $c\bar{c}s\bar{s}$  &$7.92\pm1.95$ &$7.05\pm1.45$\\ \hline
            $c\bar{c}q\bar{q} $&$6.32\pm2.30$ & $5.85\pm1.30$\\      \hline
    $b\bar{b}s\bar{s}$   & $4.46\pm1.04$&$3.68\pm0.80$\\ \hline
           $ b\bar{b}q\bar{q} $ & $3.35\pm1.00$&$3.06\pm0.66$\\      \hline
    \hline
\end{tabular}
\end{center}
\caption{ The  pole residues (in unit of $10^{-2}\, \rm{GeV}^5$ and
$10^{-1}\, \rm{GeV}^5$ for the hidden charm and hidden  bottom
channels respectively) of the scalar tetraquark states. }
\end{table}

The $SU(3)$ breaking effects for the masses of the hidden charm and
hidden bottom tetraquark states are buried in the uncertainties. The
$C\gamma_\mu-C\gamma^\mu$ type and the $C\gamma_5-C\gamma_5$ type
interpolating currents result in almost the same ground state
masses, while the ground masses of the
$C\gamma_\mu\gamma_5-C\gamma^\mu\gamma_5$ type tetraquark states are
larger than the corresponding ones of the $C\gamma_\mu-C\gamma^\mu$
type tetraquark states about $(0.2-0.5)\,\rm{GeV}$. Naively, we
expect the axial and vector diquarks have larger masses than the
corresponding scalar diqaurks, and the $C\gamma_\mu-C\gamma^\mu$
type and the $C\gamma_\mu\gamma_5-C\gamma^\mu\gamma_5$ type scalar
tetraquark states have larger masses than the corresponding
$C\gamma_5-C\gamma_5$ type scalar tetraquark states, because the
attractive interactions of one-gluon exchange favor formation of the
diquarks in  color antitriplet $\overline{3}_{ c}$, flavor
antitriplet $\overline{3}_{ f}$ and spin singlet $1_s$
\cite{GI1,GI2}.

The  meson  $Z(4250)$   may be a scalar tetraquark state
($c\bar{c}u\bar{d}$), irrespective of the $C\gamma_\mu-C\gamma^\mu$
type and the $C\gamma_5-C\gamma_5$ type diquark structures
\cite{Wang08072,WangScalar}, the decay $ Z(4250) \to \pi^+\chi_{c1}$
can take place with the OZI super-allowed "fall-apart" mechanism,
which can take into account the large total width naturally.   Other
possibilities, such as a hadro-charmonium resonance and a
$D_1^+\bar{D}^0+ D^+\bar{D}_1^0$ molecular state are not excluded;
more experimental data are still needed to identify it. It is
difficult to identify  the $Z(4050)$ as the  scalar tetraquark state
as the lower bound of   the $C\gamma_\mu-C\gamma^\mu$ type and the
$C\gamma_5-C\gamma_5$ type scalar tetraquark states are larger than
the $Z(4050)$ about $130\,\rm{MeV}$.

In this article, we calculate the mass spectrum of the scalar hidden
charm and hidden bottom tetraquark states consist of the
$C\gamma_\mu \gamma_5-C\gamma^\mu \gamma_5$ type and the
$C\gamma_\mu -C\gamma^\mu $ type diquark pairs by imposing the two
criteria of the QCD sum rules. In fact, we usually consult the
experimental data in choosing the Borel parameter $M^2$ and the
threshold parameter $s_0$ \cite{ColangeloReview}. There lack
experimental data for the phenomenological hadronic spectral
densities of the tetraquark states, the present predictions can't be
confronted with the experimental data.  The nonet scalar mesons
below $1\,\rm{GeV}$ (the $f_0(980)$ and $a_0(980)$ especially) are
good candidates for the tetraquark states
\cite{Jaffe2004,Close2002,ReviewScalar}. However, they can't satisfy
the two criteria of the QCD sum rules, and result in a reasonable
Borel window, although it is not an indication non-existence of the
light tetraquark states (For detailed discussions about this
subject, one can consult Refs.\cite{Wang08072,Wang0708}). The QCD
sum rules is just a QCD model.

For the conventional  mesons and  baryons, the Borel window
$M^2_{max}-M^2_{min}$  is rather large and  reliable QCD sum rules
can be obtained. However, for the multiquark states i.e. tetraquark
states, pentaquark states, hexaquark states, etc, the spectral
densities $\rho\sim s^n$ with $n$ is larger than the
 ones for the conventional hadrons,     integral
$\int_0^{\infty} s^n \exp \{-\frac{s}{M^2}\} ds$ converges more
slowly, which results in some sensitivities  to the threshold
parameters inevitably.  We select the threshold parameters and Borel
parameters by imposing   the two criteria of the QCD sum rules, and
intend to select the possibly  lowest threshold parameter which
corresponds  to the ground state.

In Table 1, we also present the results for the
$C\gamma_\mu-C\gamma^\mu$ type scalar tetraquark states from a
relativistic quark model based on a quasipotential approach in QCD
\cite{Ebert0512,Ebert0812}, the central values of our predictions
are larger than the corresponding ones from the quasipotential model
about $(0.1-0.5)\, \rm{GeV}$. The predications based on constituent
diquark model ($M_{c\bar{c}q\bar{q}}=3723\,\rm{MeV}$
 \cite{Maiani20042} and $M_{c\bar{c}s\bar{s}}=3834\,\rm{MeV}$ \cite{Polosa0902} for the
 tetraquark states $c\bar{c}q\bar{q}$   and $c\bar{c}s\bar{s}$
 respectively) are about $0.6\,\rm{GeV}$ smaller than the corresponding ones in the present work.

The predictions of Refs.\cite {Maiani20042,Maiani2008,Polosa0902}
depend heavily on the assumption that the light
 scalar mesons $a_0(980)$ and $f_0(980)$ are tetraquark states,
 the  basic  parameters (constituent diquark masses) are
 estimated thereafter.
In the conventional quark models, the constituent quark masses  are
taken as the basic input parameters, and fitted to reproduce the
mass spectra  of the well known  mesons and baryons. However, the
present experimental knowledge about the phenomenological hadronic
spectral densities of the tetraquark states is  rather vague,
whether or not there exist   tetraquark states is not confirmed with
confidence. The predicted constituent diquark masses cannot be
confronted with the experimental data.

The LHCb is a dedicated $b$ and $c$-physics precision experiment at
the LHC (large hadron collider). The LHC will be the world's most
copious  source of the $b$ hadrons, and  a complete spectrum of the
$b$ hadrons will be available through gluon fusion. In proton-proton
collisions at $\sqrt{s}=14\,\rm{TeV}$¡Ì, the $b\bar{b}$ cross
section is expected to be $\sim 500\mu b$ producing $10^{12}$
$b\bar{b}$ pairs in a standard  year of running at the LHCb
operational luminosity of $2\times10^{32} \rm{cm}^{-2}
\rm{sec}^{-1}$ \cite{LHC}. The scalar tetraquark states
(irrespective of the $C\gamma_\mu-C\gamma^\mu$ type,  the
$C\gamma_\mu\gamma_5-C\gamma^\mu\gamma_5$ type  and the
$C\gamma_5-C\gamma_5$ type diquark structures)  may be observed at
the LHCb, if they exist indeed. We can search for the scalar hidden
charm tetraquark states in the $D\bar{D}$, $D^*\bar{D^*}$,
$D_s\bar{D_s}$, $D_s^*\bar{D_s^*}$, $J/\psi \rho$, $J/\psi \phi$,
$J/\psi \omega$, $\eta_c\pi$, $\eta_c\eta$, $\cdots$ invariant mass
distributions and search for the scalar hidden bottom tetraquark
states  in the $B\bar{B}$, $B^*\bar{B^*}$, $B_s\bar{B_s}$,
$B_s^*\bar{B_s^*}$, $\Upsilon \rho$, $\Upsilon \phi$, $\Upsilon
\omega$, $\eta_b\pi$, $\eta_b\eta$, $\cdots$ invariant mass
distributions.

\begin{figure}
 \centering
 \includegraphics[totalheight=4cm,width=5cm]{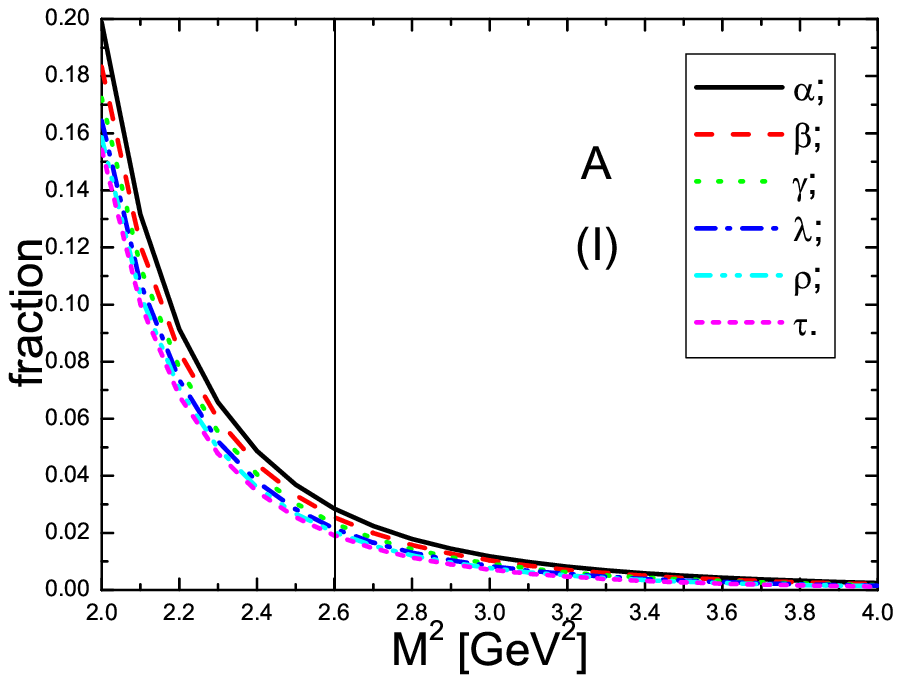}
 \includegraphics[totalheight=4cm,width=5cm]{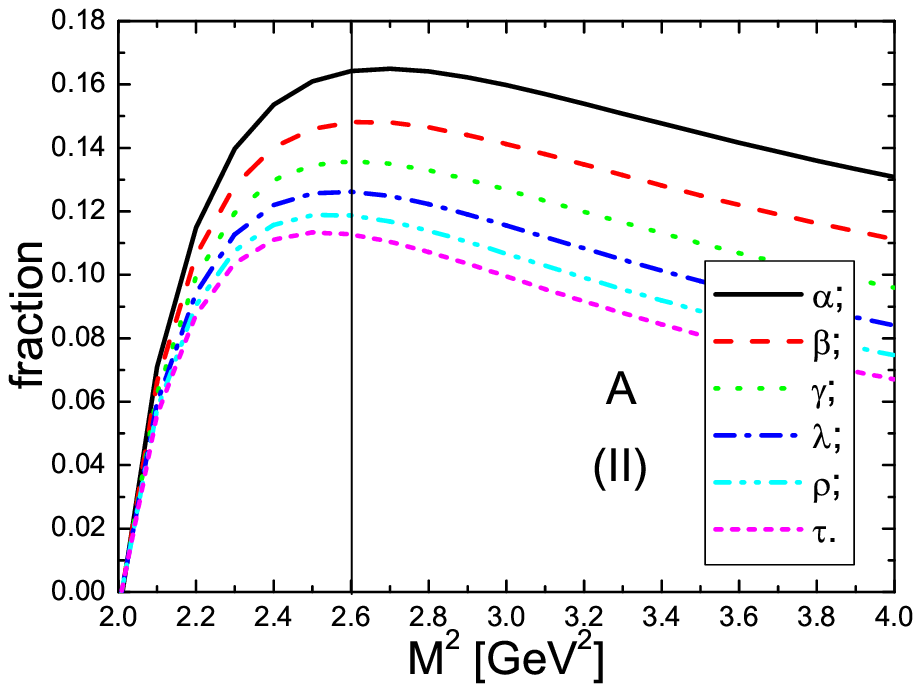}
 \includegraphics[totalheight=4cm,width=5cm]{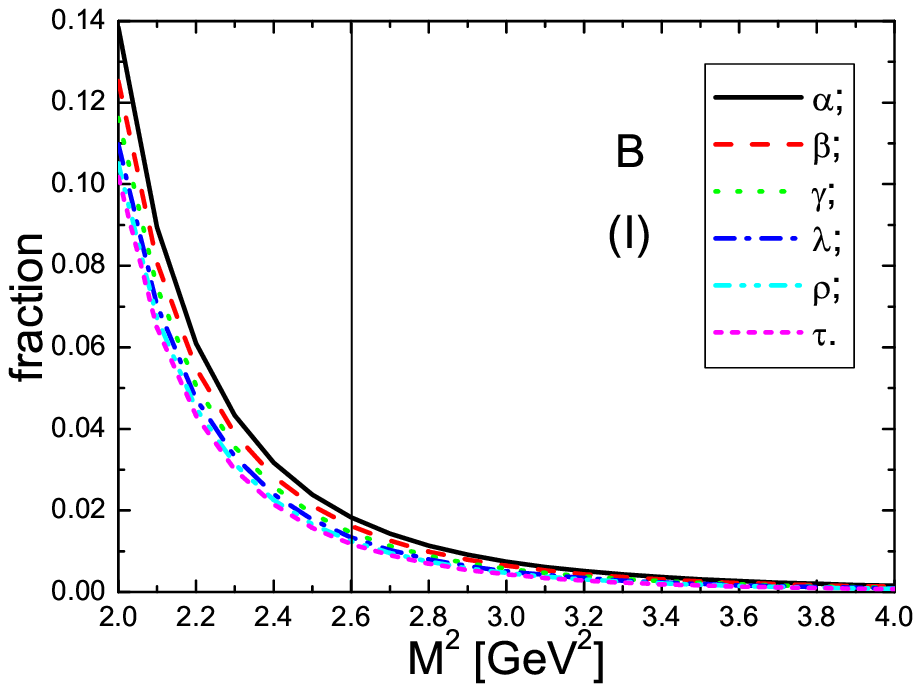}
 \includegraphics[totalheight=4cm,width=5cm]{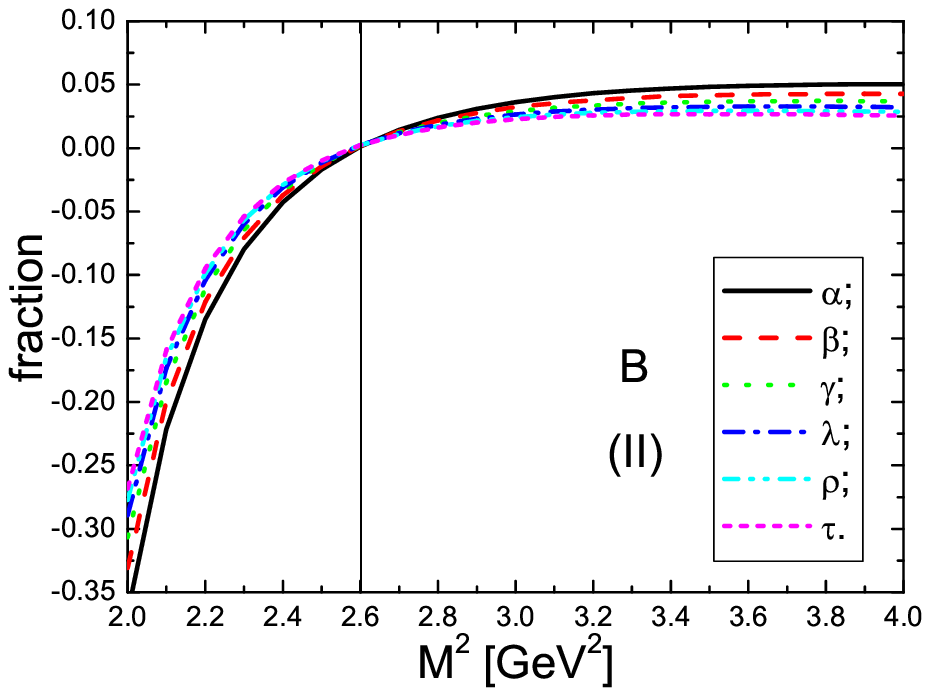}
\includegraphics[totalheight=4cm,width=5cm]{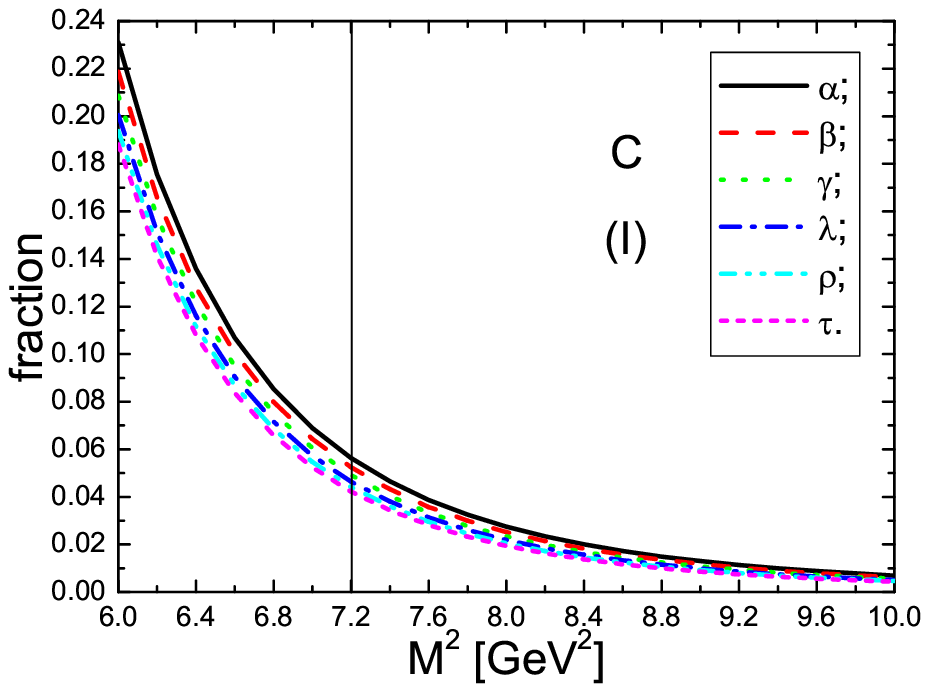}
\includegraphics[totalheight=4cm,width=5cm]{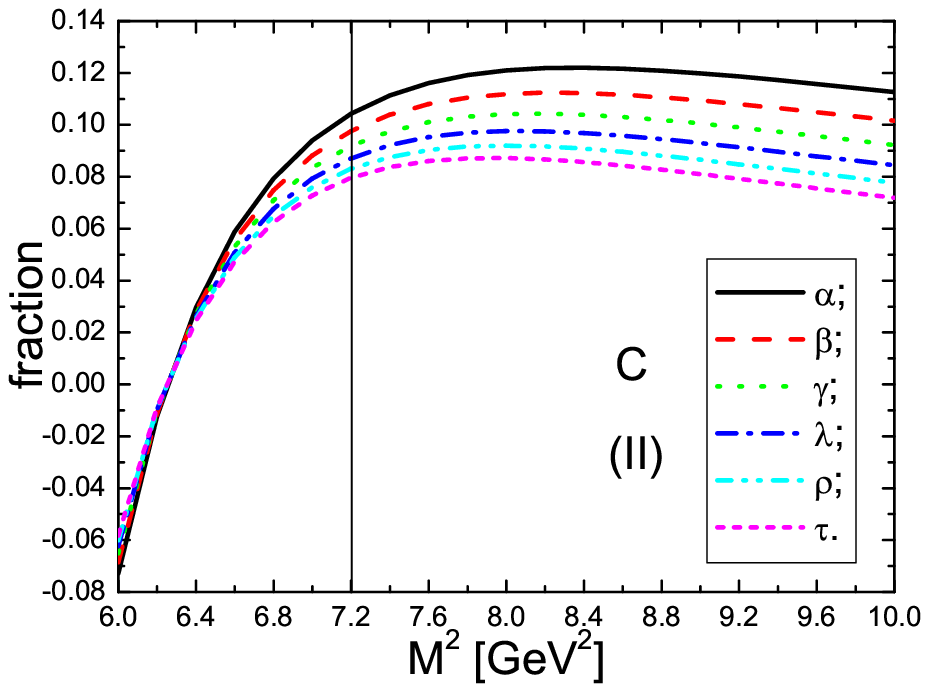}
\includegraphics[totalheight=4cm,width=5cm]{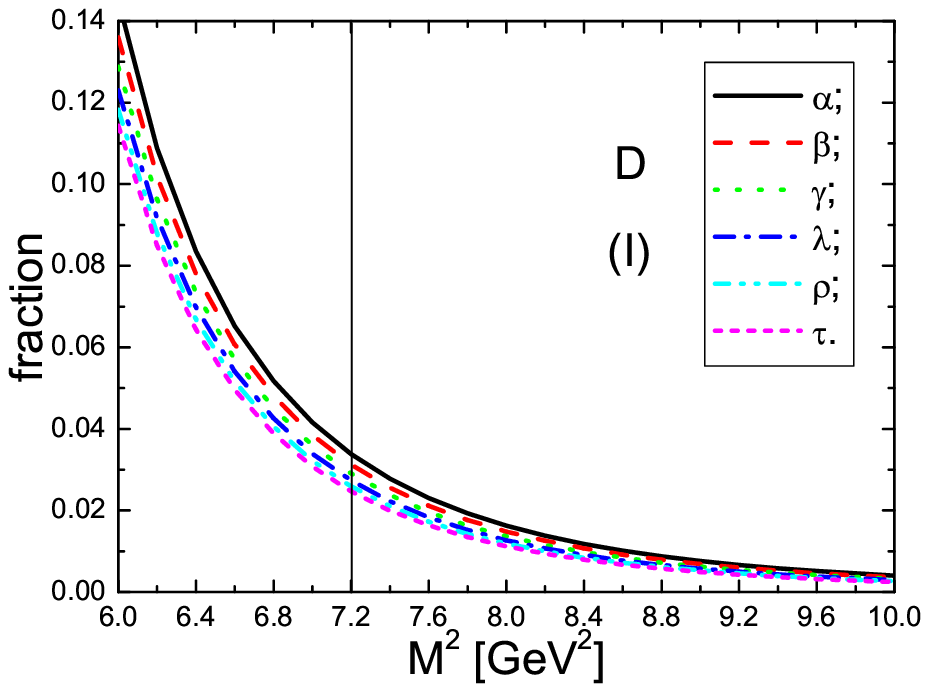}
 \includegraphics[totalheight=4cm,width=5cm]{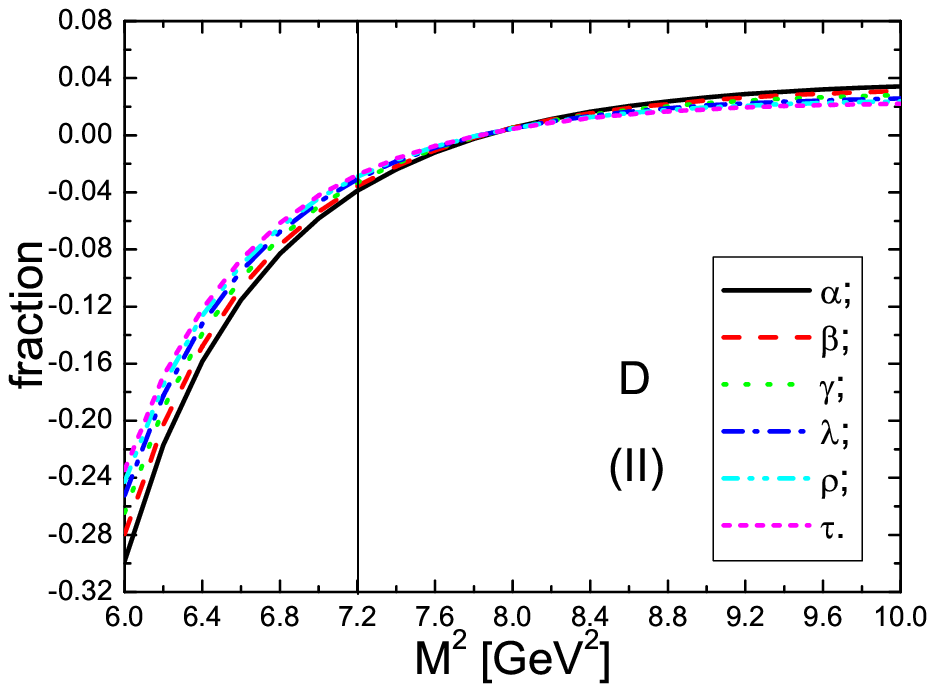}
     \caption{ The contributions from  the high dimension vacuum condensates  with variation of the Borel
   parameter $M^2$  in the operator product expansion for the $C\gamma_\mu-C\gamma^\mu$ type interpolating currents.
   The   (I) and (II)  denote the  contributions from the $\langle \bar{q}g_s\sigma G q \rangle^2$
   and     $\langle \bar{q} q \rangle^2$ +$\langle \bar{q} q \rangle\langle \bar{q}g_s\sigma G q \rangle$ terms respectively.
      The $A$,  $B$, $C$ and $D$  correspond to
   the $c\bar{c}q\bar{q} $, $c\bar{c}s\bar{s} $, $b\bar{b}q\bar{q} $ and $b\bar{b}s\bar{s} $ channels respectively. The
notations   $\alpha$, $\beta$, $\gamma$, $\lambda$, $\rho$ and
$\tau$  correspond to the threshold
   parameters $s_0=21\,\rm{GeV}^2$,
   $22\,\rm{GeV}^2$, $23\,\rm{GeV}^2$, $24\,\rm{GeV}^2$, $25\,\rm{GeV}^2$ and $26\,\rm{GeV}^2$ respectively in the hidden charm channels;
   while in the hidden bottom channels they correspond to the threshold
   parameters $s_0=132\,\rm{GeV}^2$,
   $134\,\rm{GeV}^2$, $136\,\rm{GeV}^2$, $138\,\rm{GeV}^2$, $140\,\rm{GeV}^2$ and $142\,\rm{GeV}^2$ respectively.}
\end{figure}

\begin{figure}
 \centering
 \includegraphics[totalheight=4cm,width=5cm]{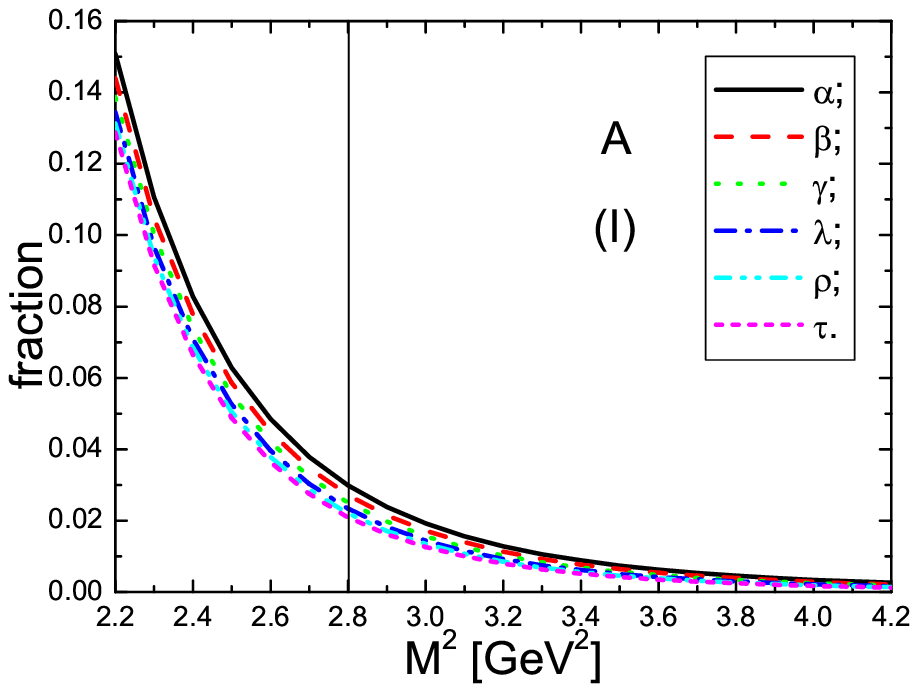}
 \includegraphics[totalheight=4cm,width=5cm]{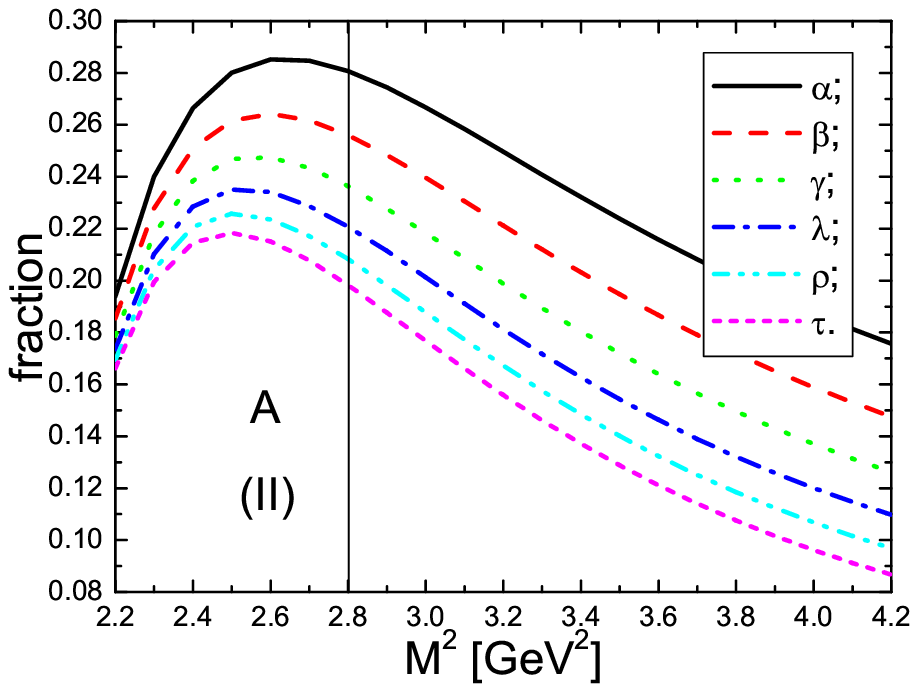}
 \includegraphics[totalheight=4cm,width=5cm]{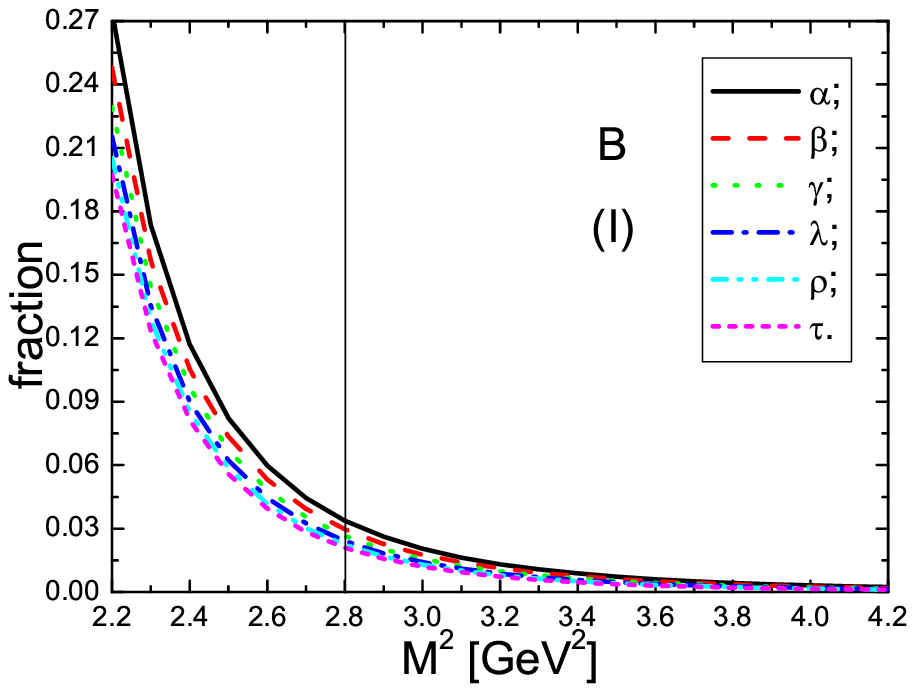}
 \includegraphics[totalheight=4cm,width=5cm]{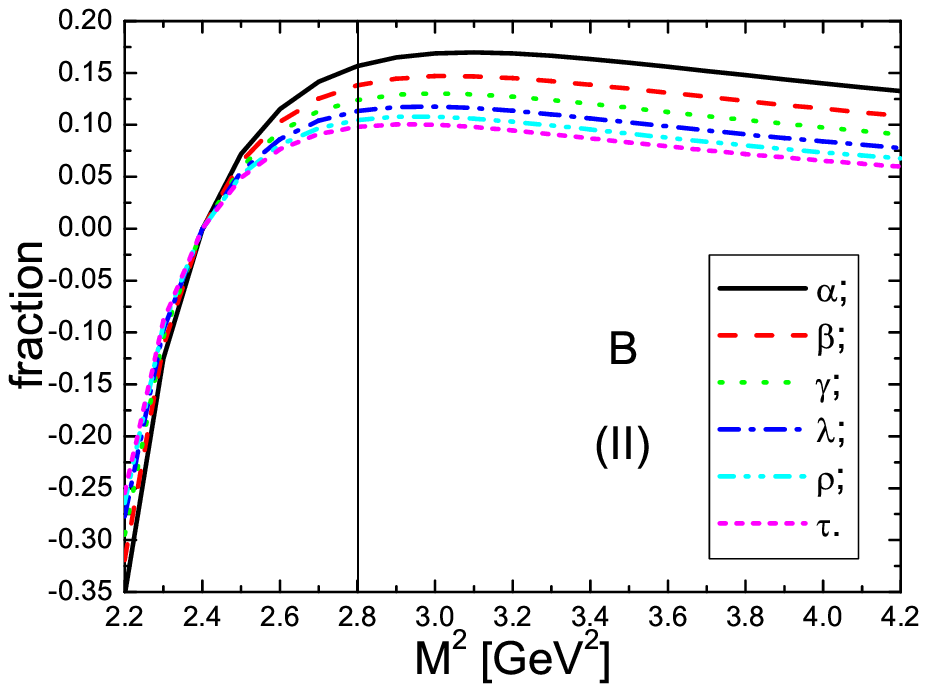}
\includegraphics[totalheight=4cm,width=5cm]{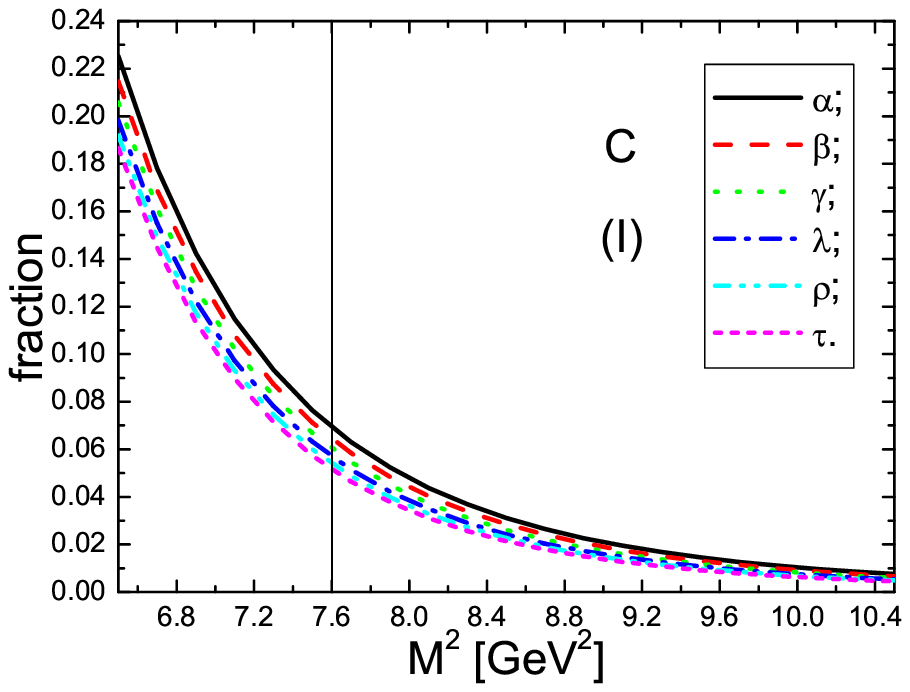}
\includegraphics[totalheight=4cm,width=5cm]{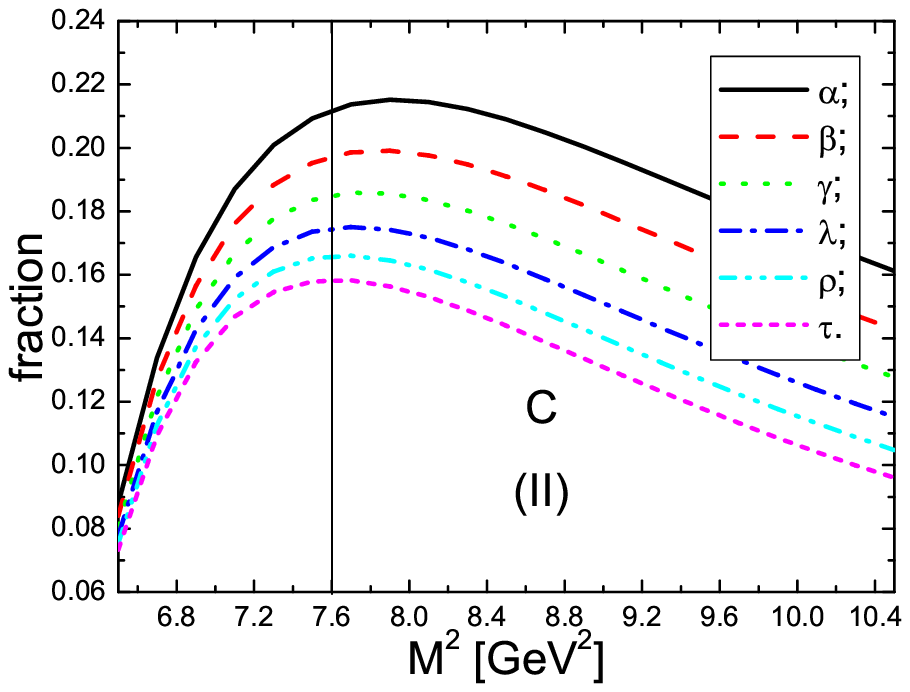}
\includegraphics[totalheight=4cm,width=5cm]{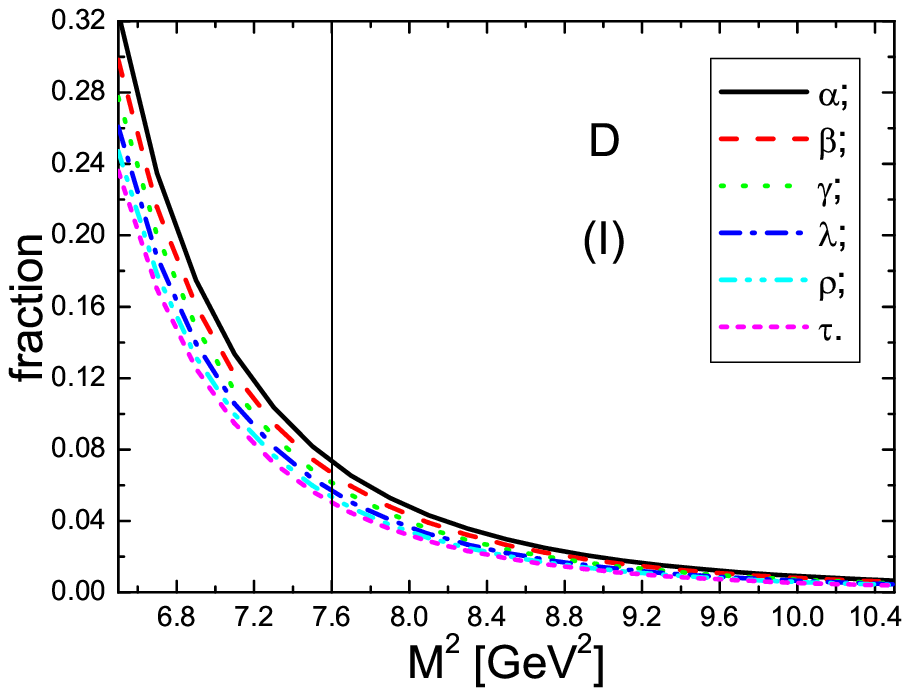}
 \includegraphics[totalheight=4cm,width=5cm]{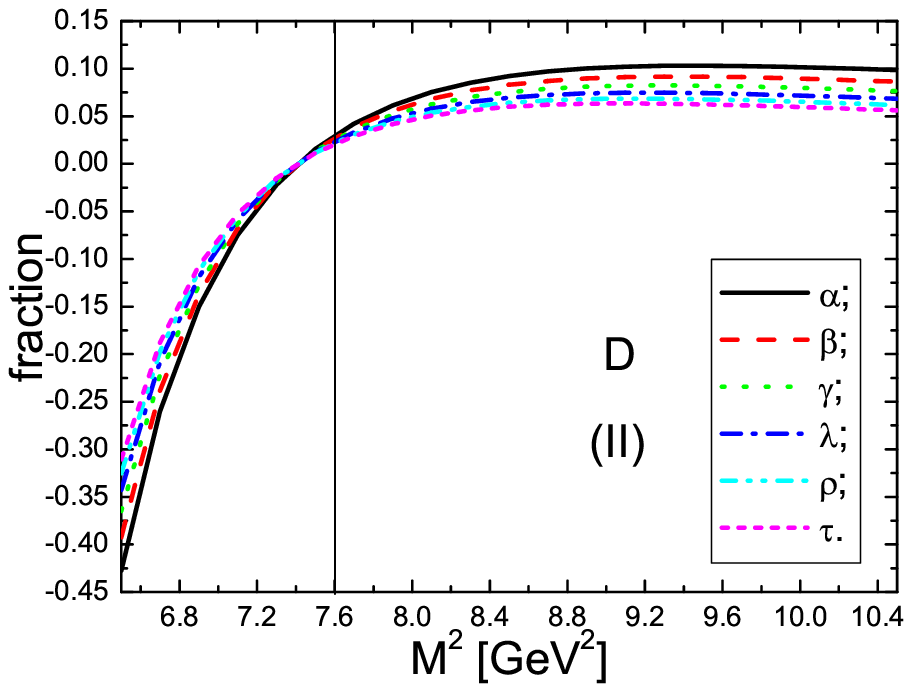}
     \caption{ The contributions from  the high dimension vacuum condensates  with variation of the Borel
   parameter $M^2$  in the operator product expansion for the $C\gamma_\mu\gamma_5-C\gamma^\mu\gamma_5$ type interpolating currents.
   The   (I) and (II)  denote the  contributions from the $\langle \bar{q}g_s\sigma G q \rangle^2$
   and     $\langle \bar{q} q \rangle^2$ +$\langle \bar{q} q \rangle\langle \bar{q}g_s\sigma G q \rangle$ terms respectively.
      The $A$,  $B$, $C$ and $D$  correspond to
   the $c\bar{c}q\bar{q} $, $c\bar{c}s\bar{s} $, $b\bar{b}q\bar{q} $ and $b\bar{b}s\bar{s} $ channels respectively. The
notations   $\alpha$, $\beta$, $\gamma$, $\lambda$, $\rho$ and
$\tau$  correspond to the threshold
   parameters $s_0=25\,\rm{GeV}^2$,
   $26\,\rm{GeV}^2$, $27\,\rm{GeV}^2$, $28\,\rm{GeV}^2$, $29\,\rm{GeV}^2$ and $30\,\rm{GeV}^2$ respectively in the hidden charm channels;
   while in the hidden bottom channels they correspond to the threshold
   parameters $s_0=142\,\rm{GeV}^2$,
   $144\,\rm{GeV}^2$, $146\,\rm{GeV}^2$, $148\,\rm{GeV}^2$, $150\,\rm{GeV}^2$ and $152\,\rm{GeV}^2$ respectively.}
\end{figure}

\begin{figure}
 \centering
 \includegraphics[totalheight=4cm,width=5cm]{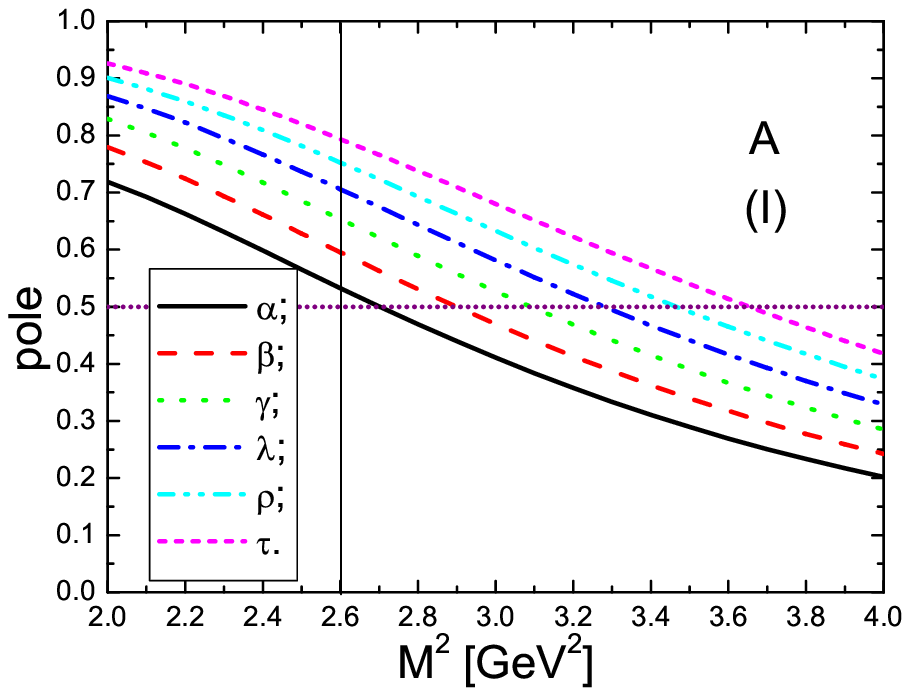}
 \includegraphics[totalheight=4cm,width=5cm]{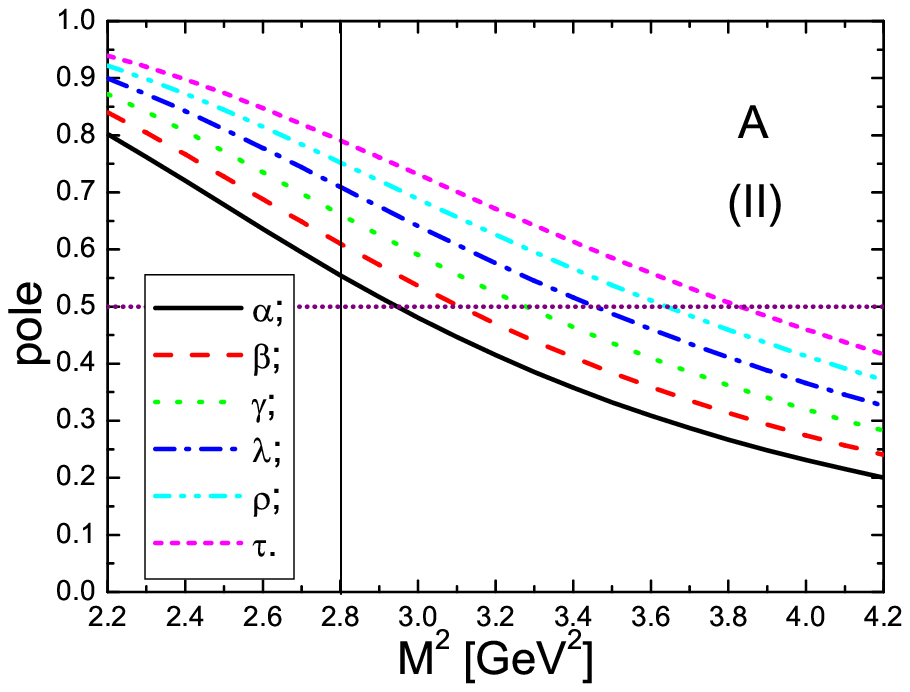}
 \includegraphics[totalheight=4cm,width=5cm]{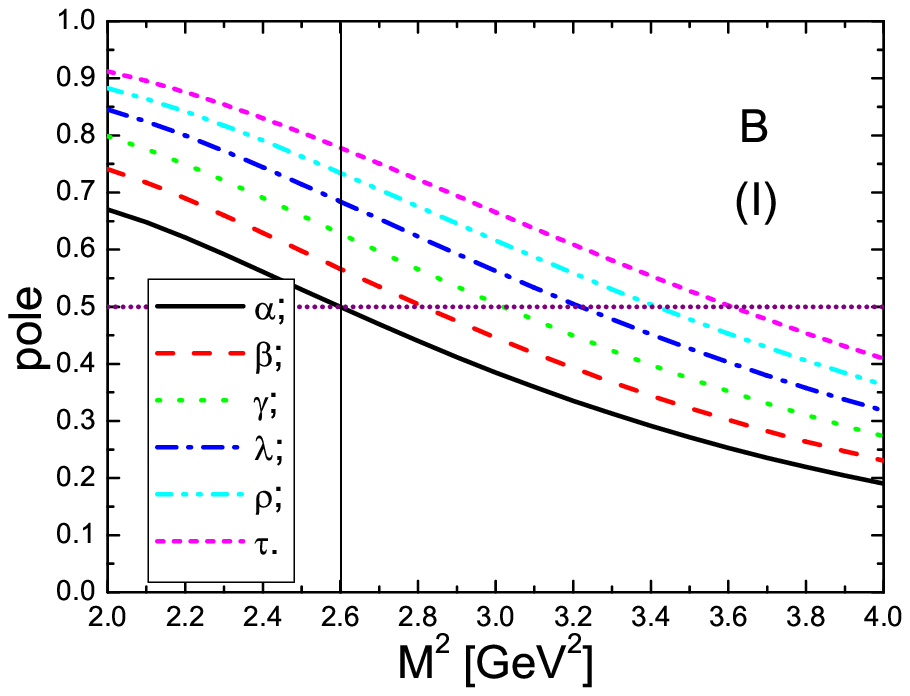}
  \includegraphics[totalheight=4cm,width=5cm]{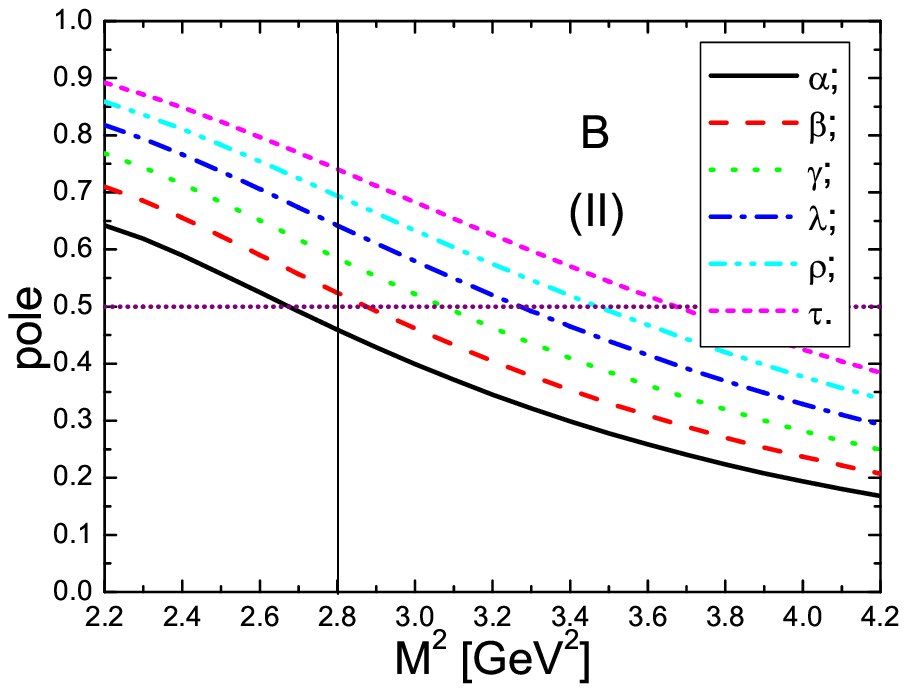}
 \includegraphics[totalheight=4cm,width=5cm]{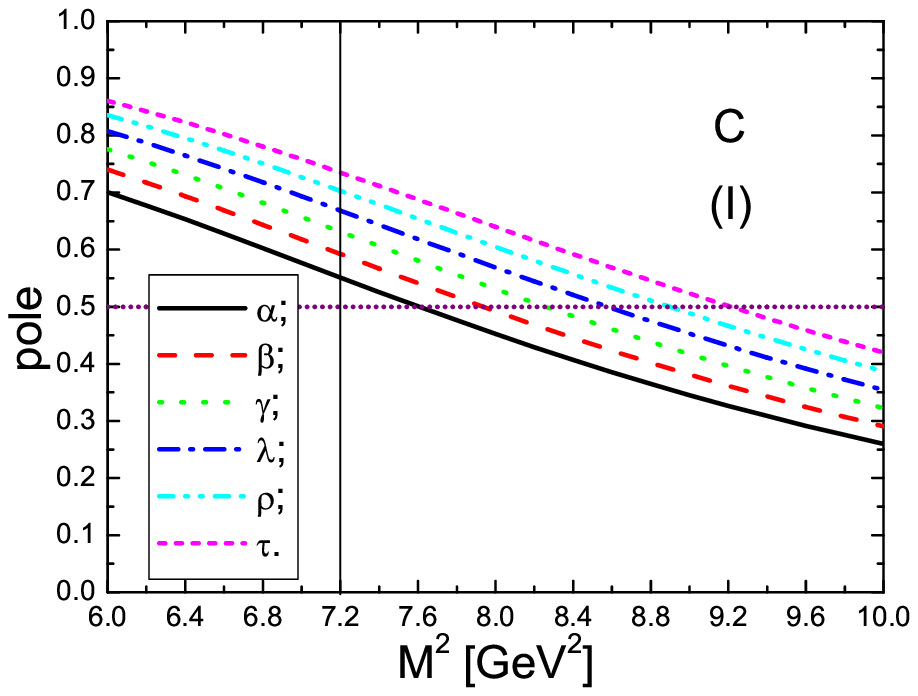}
 \includegraphics[totalheight=4cm,width=5cm]{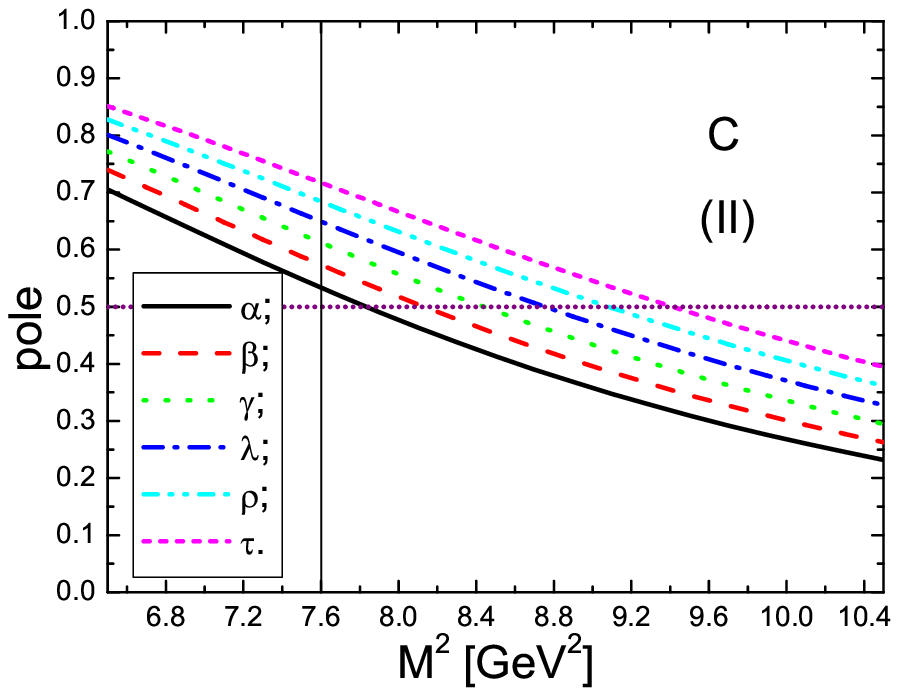}
\includegraphics[totalheight=4cm,width=5cm]{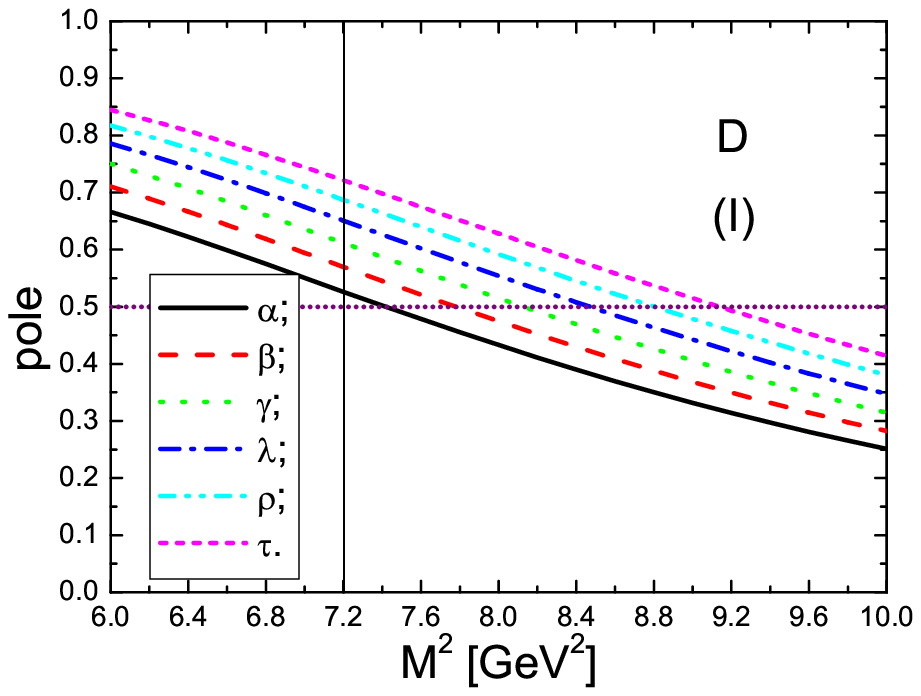}
 \includegraphics[totalheight=4cm,width=5cm]{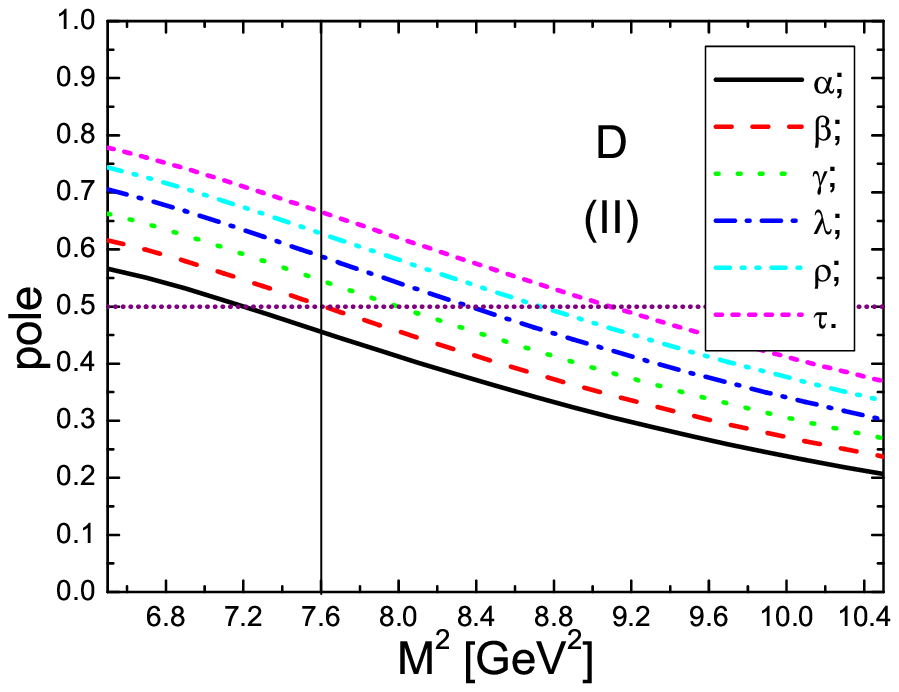}
   \caption{ The contributions from the pole terms with variation of the Borel parameter $M^2$. The $A$, $B$, $C$,
   and $D$ denote the $c\bar{c}q\bar{q}$,
    $c\bar{c}s\bar{s}$, $b\bar{b}q\bar{q}$,
   and $b\bar{b}s\bar{s}$ channels respectively. The (I) and (II) denote the $C\gamma_\mu-C\gamma^\mu$ type and
   the $C\gamma_\mu\gamma_5-C\gamma^\mu\gamma_5$
   type interpolating currents respectively. For the
   $C\gamma_\mu-C\gamma^\mu$ ($C\gamma_\mu\gamma_5-C\gamma^\mu\gamma_5$)
   type interpolating currents,  in the hidden charm channels, the notations
   $\alpha$, $\beta$, $\gamma$, $\lambda$, $\rho$ and $\tau$  correspond to the threshold
   parameters $s_0=21\,\rm{GeV}^2 \,(25\,\rm{GeV}^2)$,
   $22\,\rm{GeV}^2\,(26\,\rm{GeV}^2)$, $23\,\rm{GeV}^2\,(27\,\rm{GeV}^2)$,
   $24\,\rm{GeV}^2\,(28\,\rm{GeV}^2)$, $25\,\rm{GeV}^2\,(29\,\rm{GeV}^2)$
   and $26\,\rm{GeV}^2\,(30\,\rm{GeV}^2)$, respectively;  while in the hidden bottom channels they correspond to
    the threshold
   parameters  $s_0=132\,\rm{GeV}^2\,(142\,\rm{GeV}^2)$,
   $134\,\rm{GeV}^2\,(144\,\rm{GeV}^2)$, $136\,\rm{GeV}^2\,(146\,\rm{GeV}^2)$, $138\,\rm{GeV}^2\,
   (148\,\rm{GeV}^2)$, $140\,\rm{GeV}^2\,(150\,\rm{GeV}^2)$ and $142\,\rm{GeV}^2\,(152\,\rm{GeV}^2)$, respectively. }
\end{figure}

\begin{figure}
 \centering
  \includegraphics[totalheight=4cm,width=5cm]{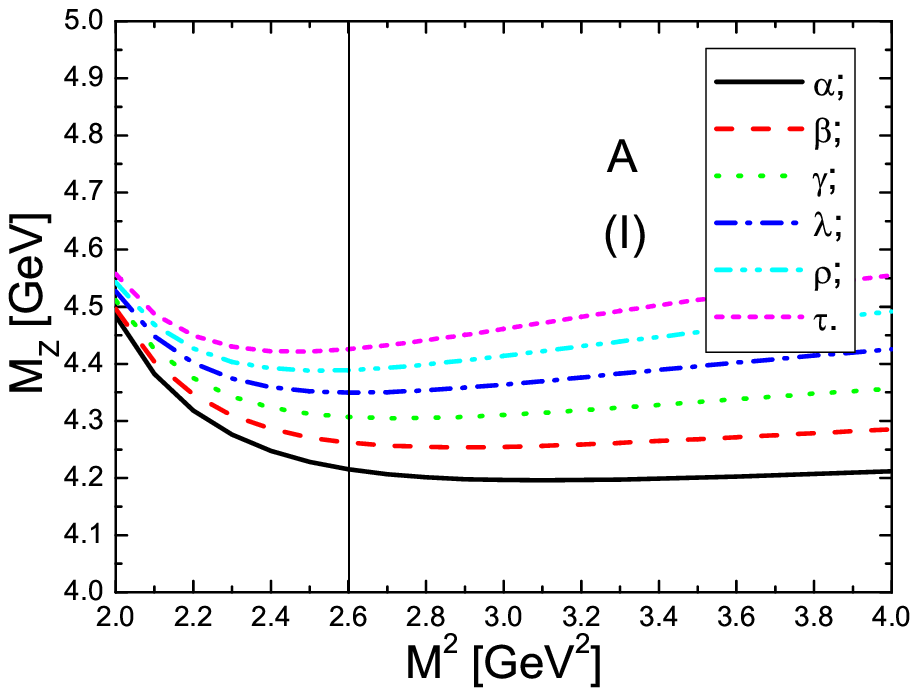}
  \includegraphics[totalheight=4cm,width=5cm]{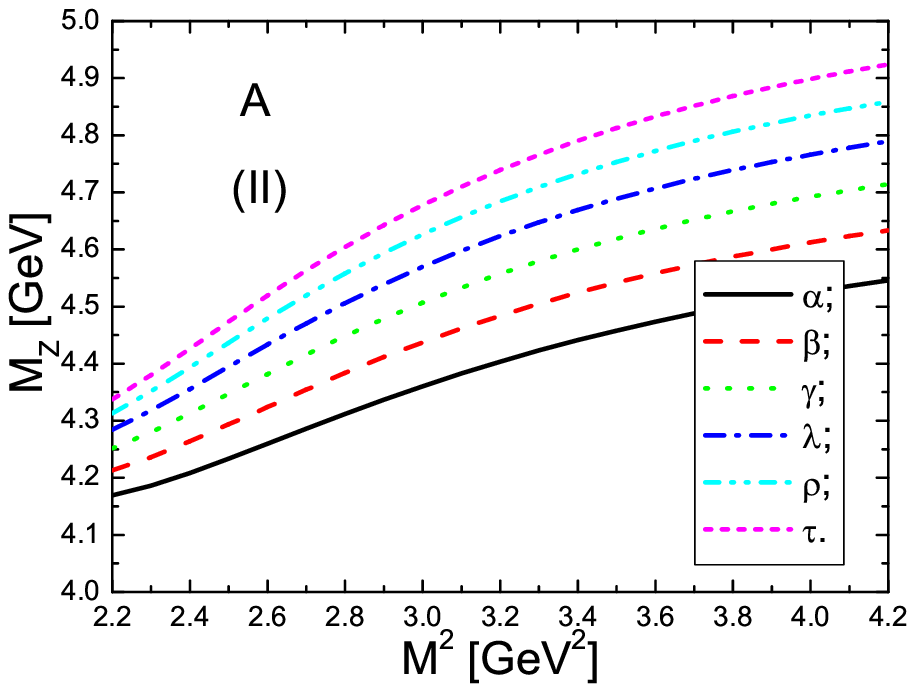}
  \includegraphics[totalheight=4cm,width=5cm]{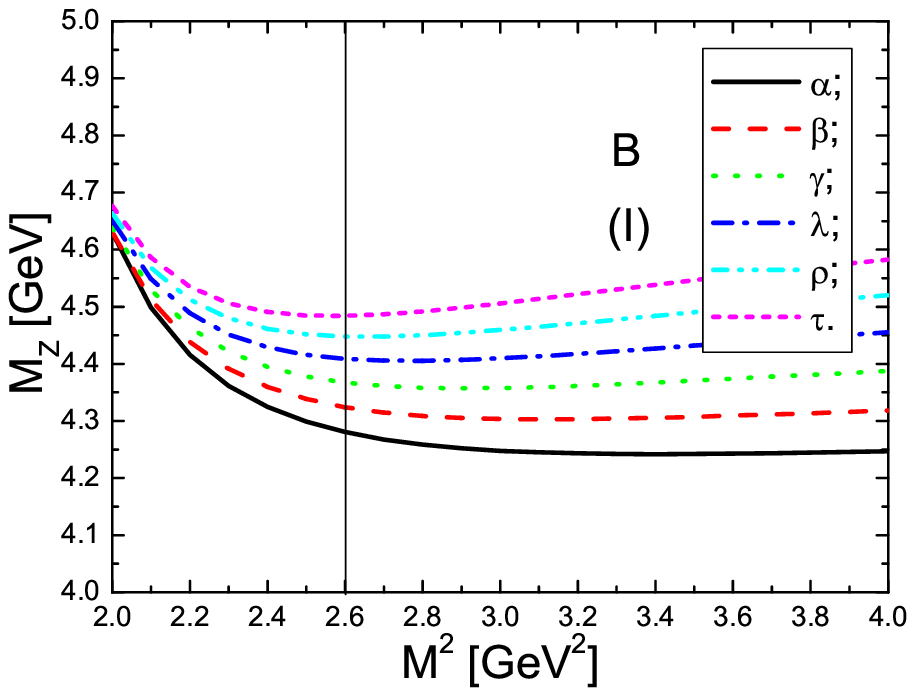}
  \includegraphics[totalheight=4cm,width=5cm]{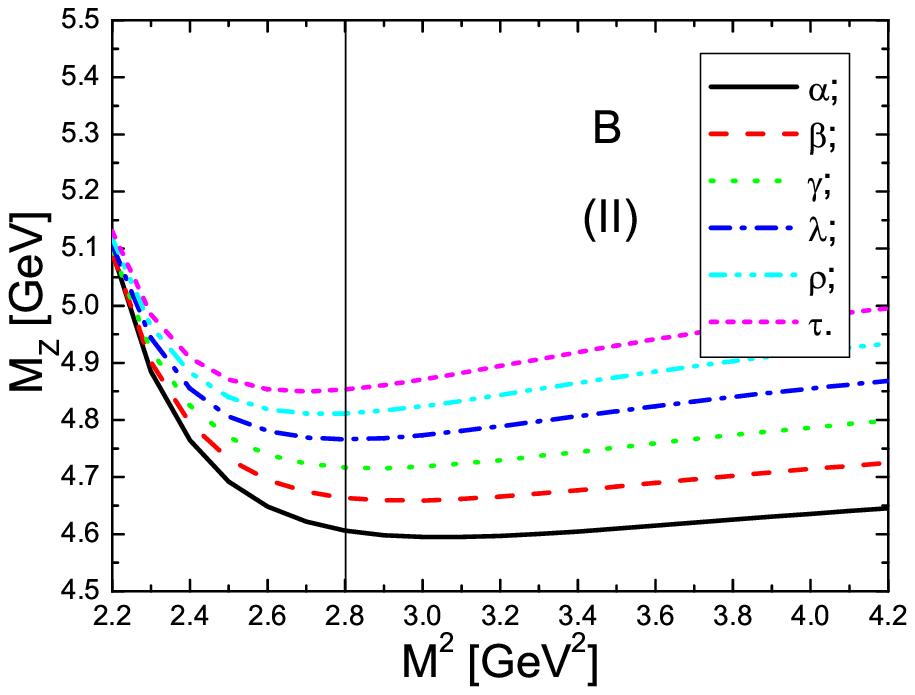}
  \includegraphics[totalheight=4cm,width=5cm]{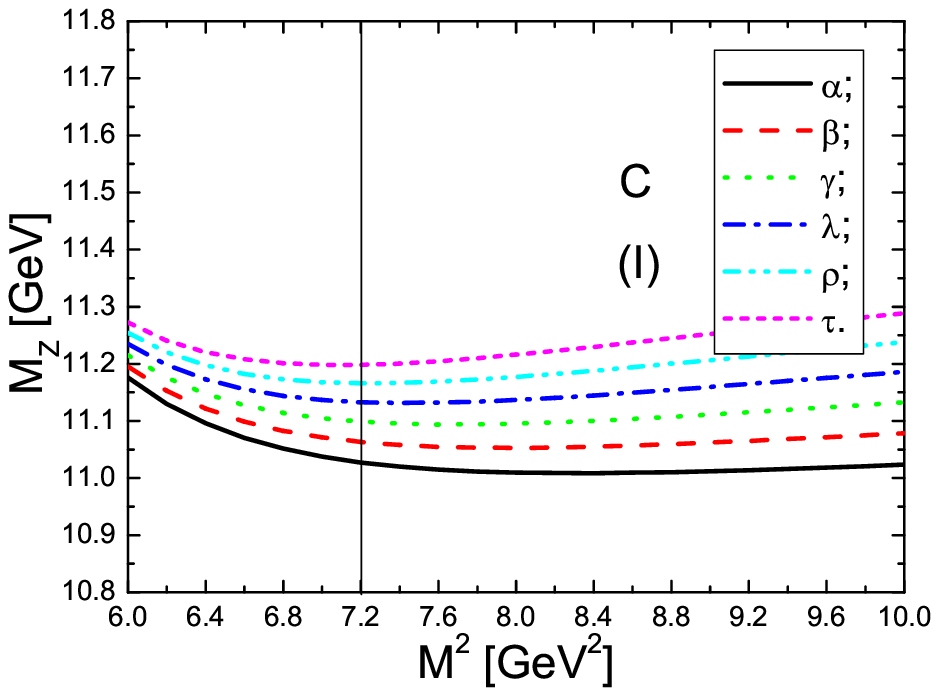}
  \includegraphics[totalheight=4cm,width=5cm]{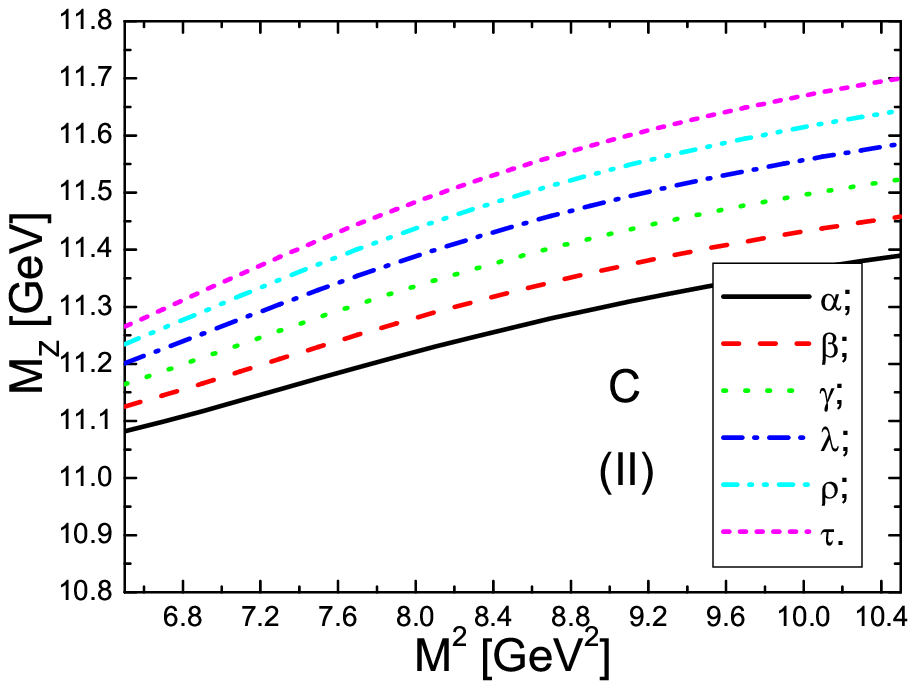}
  \includegraphics[totalheight=4cm,width=5cm]{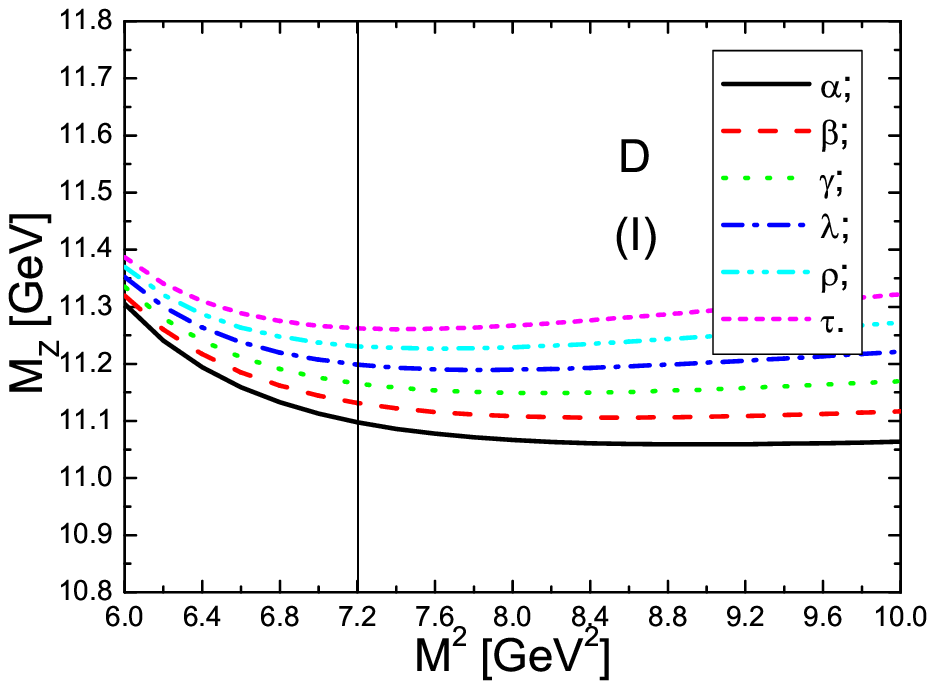}
  \includegraphics[totalheight=4cm,width=5cm]{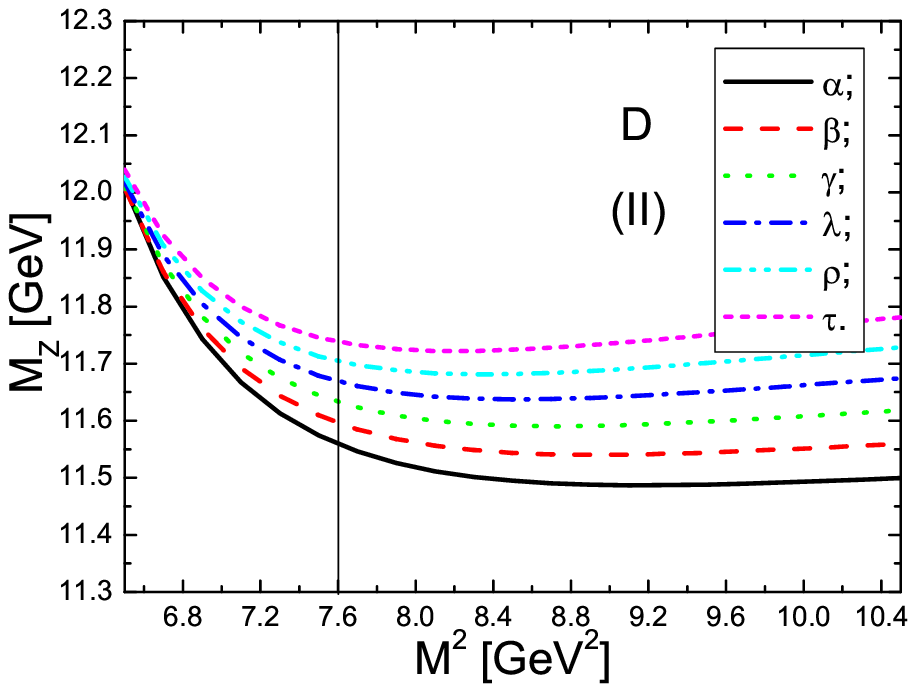}
   \caption{ The masses of the scalar tetraquark  states with variation of the Borel parameter $M^2$ and
   threshold parameter $s_0$.
    The $A$, $B$, $C$,
   and $D$ denote the $c\bar{c}q\bar{q}$,
    $c\bar{c}s\bar{s}$, $b\bar{b}q\bar{q}$,
   and $b\bar{b}s\bar{s}$ channels respectively. The (I) and (II) denote the $C\gamma_\mu-C\gamma^\mu$ type and
   the $C\gamma_\mu\gamma_5-C\gamma^\mu\gamma_5$
   type interpolating currents respectively. For the
   $C\gamma_\mu-C\gamma^\mu$ ($C\gamma_\mu\gamma_5-C\gamma^\mu\gamma_5$)
   type interpolating currents,  in the hidden charm channels, the notations
   $\alpha$, $\beta$, $\gamma$, $\lambda$, $\rho$ and $\tau$  correspond to the threshold
   parameters $s_0=21\,\rm{GeV}^2 \,(25\,\rm{GeV}^2)$,
   $22\,\rm{GeV}^2\,(26\,\rm{GeV}^2)$, $23\,\rm{GeV}^2\,(27\,\rm{GeV}^2)$,
   $24\,\rm{GeV}^2\,(28\,\rm{GeV}^2)$, $25\,\rm{GeV}^2\,(29\,\rm{GeV}^2)$
   and $26\,\rm{GeV}^2\,(30\,\rm{GeV}^2)$, respectively;  while in the hidden bottom channels they correspond to
    the threshold
   parameters  $s_0=132\,\rm{GeV}^2\,(142\,\rm{GeV}^2)$,
   $134\,\rm{GeV}^2\,(144\,\rm{GeV}^2)$, $136\,\rm{GeV}^2\,(146\,\rm{GeV}^2)$, $138\,\rm{GeV}^2\,
   (148\,\rm{GeV}^2)$, $140\,\rm{GeV}^2\,(150\,\rm{GeV}^2)$ and $142\,\rm{GeV}^2\,(152\,\rm{GeV}^2)$, respectively. }
\end{figure}

\begin{figure}
 \centering
  \includegraphics[totalheight=5cm,width=6cm]{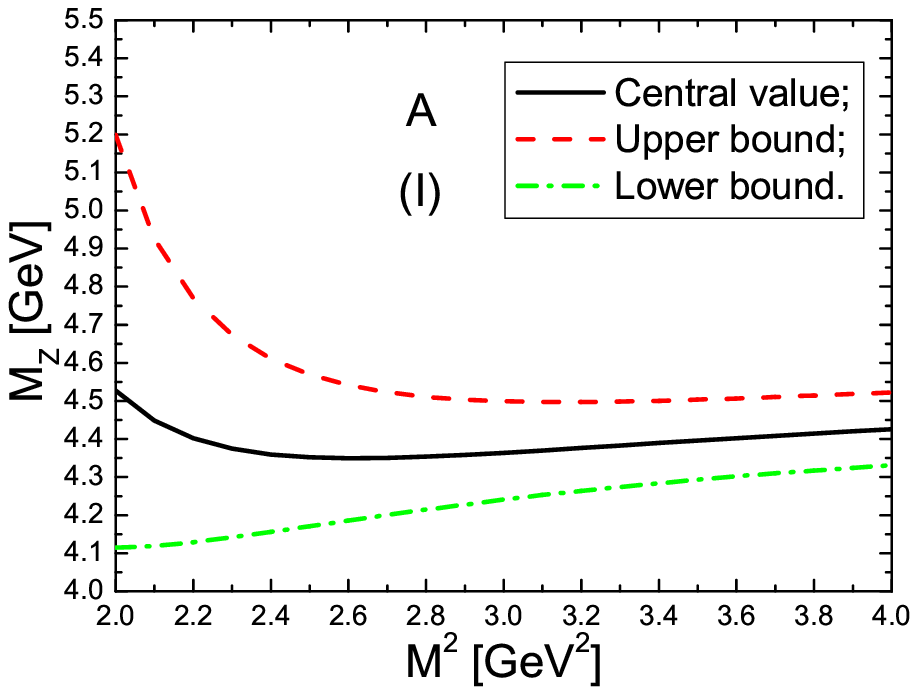}
   \includegraphics[totalheight=5cm,width=6cm]{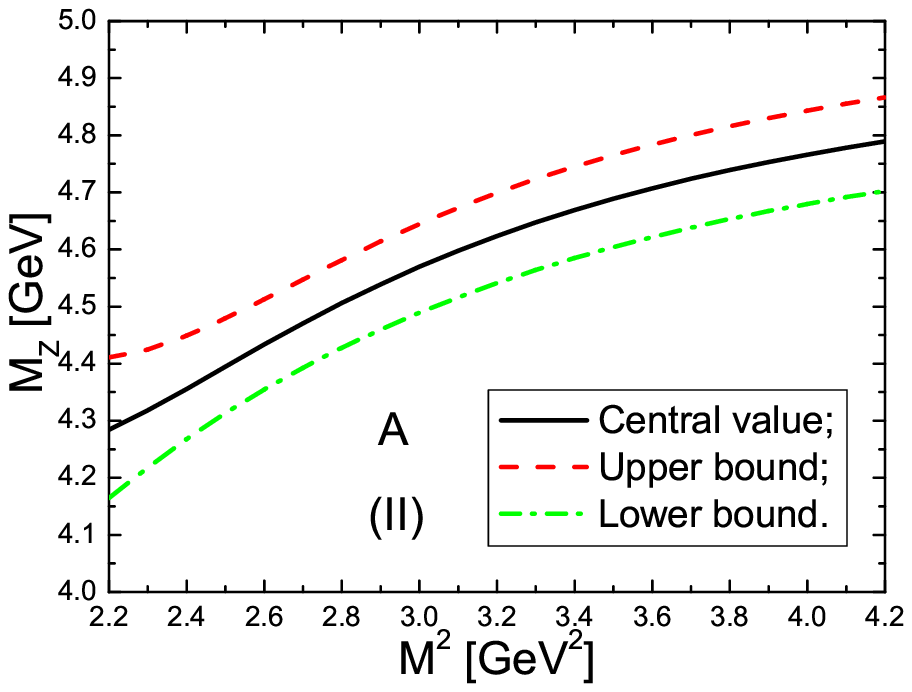}
  \includegraphics[totalheight=5cm,width=6cm]{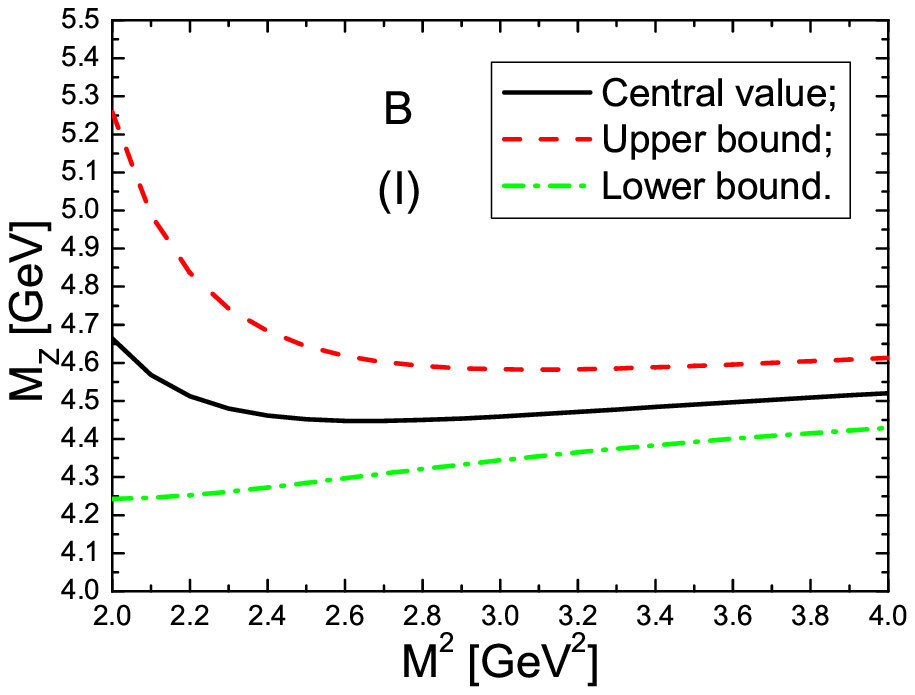}
  \includegraphics[totalheight=5cm,width=6cm]{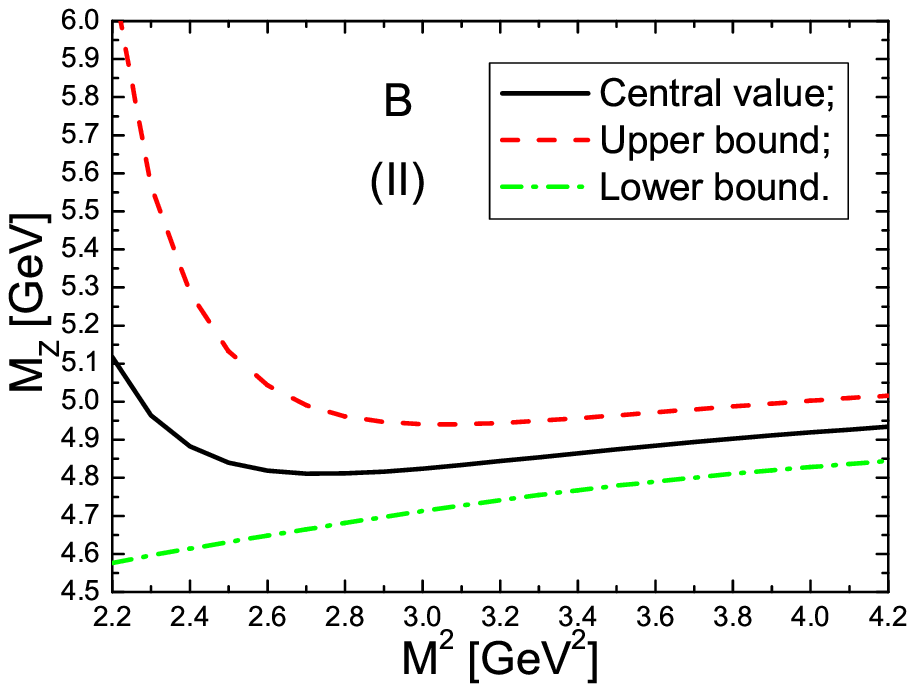}
  \includegraphics[totalheight=5cm,width=6cm]{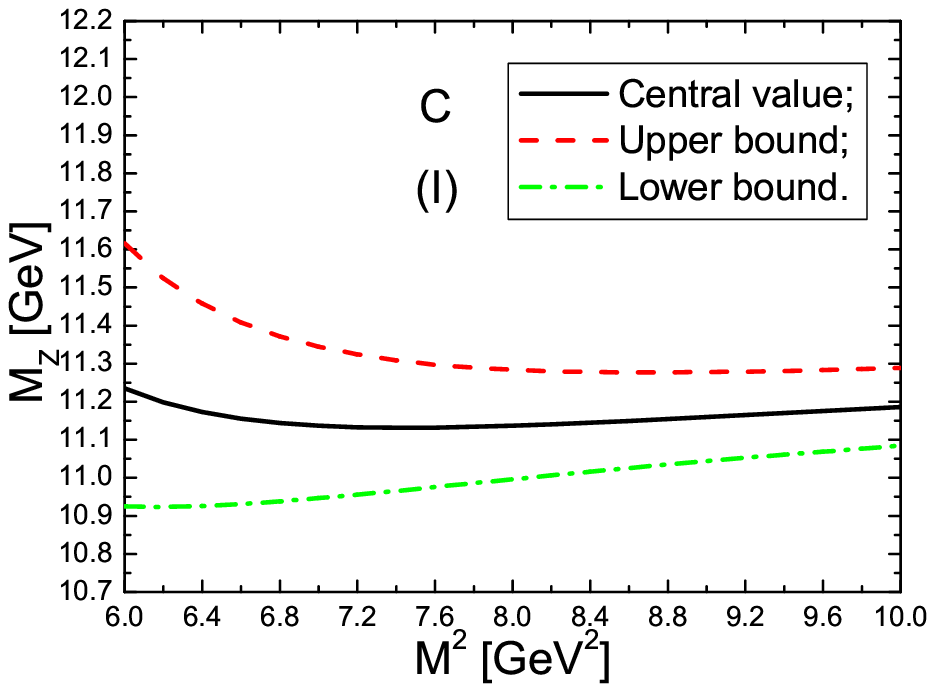}
   \includegraphics[totalheight=5cm,width=6cm]{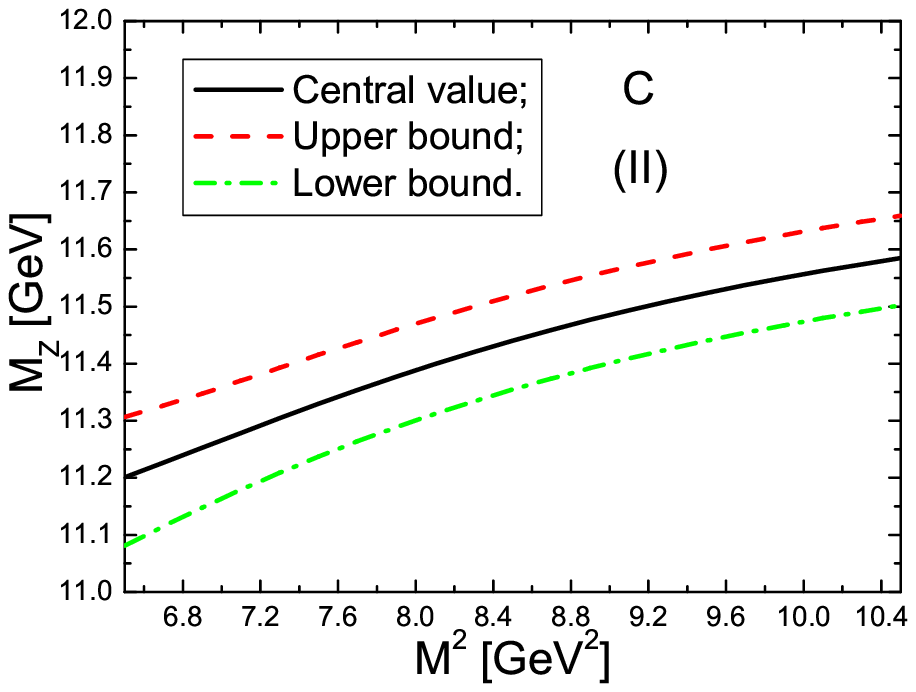}
  \includegraphics[totalheight=5cm,width=6cm]{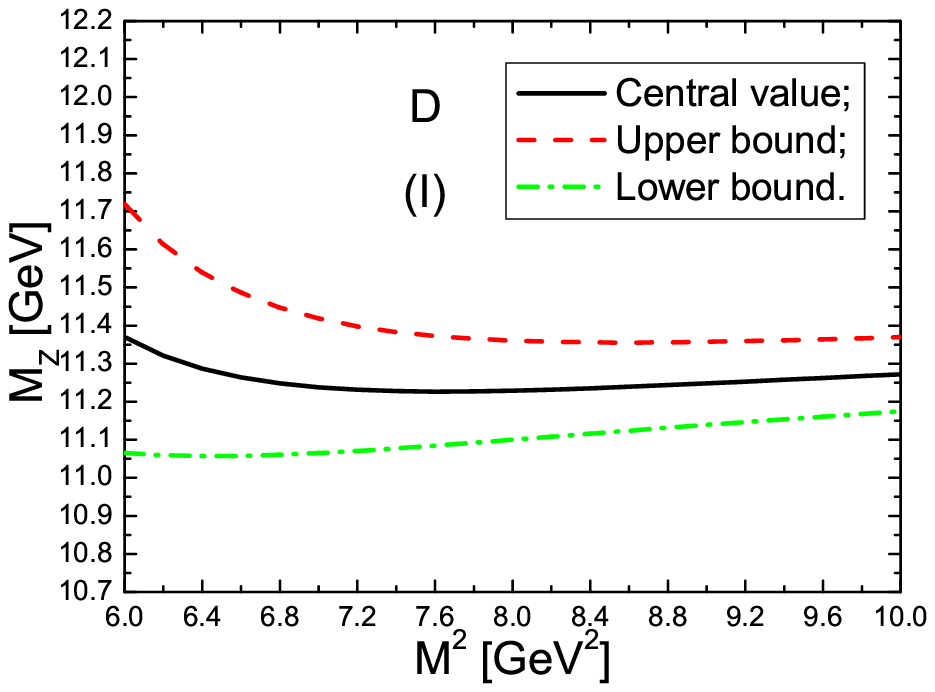}
  \includegraphics[totalheight=5cm,width=6cm]{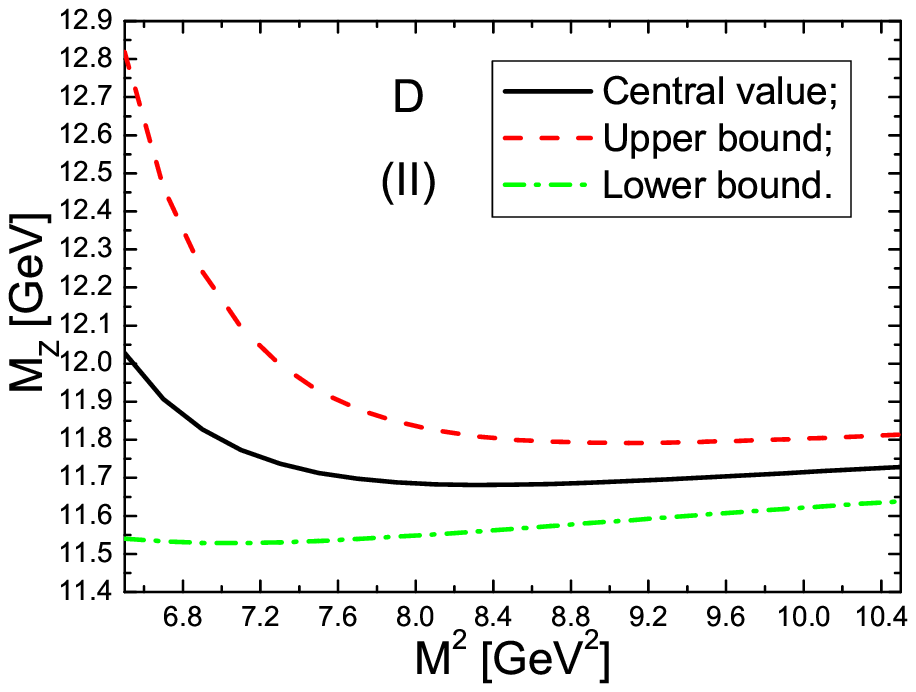}
   \caption{ The masses of the scalar tetraquark  states with variation of the Borel parameter $M^2$. The $A$, $B$, $C$,
   and $D$ denote the $c\bar{c}q\bar{q}$,
    $c\bar{c}s\bar{s}$, $b\bar{b}q\bar{q}$,
   and $b\bar{b}s\bar{s}$ channels respectively. The (I) and (II) denote the $C\gamma_\mu-C\gamma^\mu$ type
   and the
   $C\gamma_\mu\gamma_5-C\gamma^\mu\gamma_5$
   type interpolating currents respectively. }
\end{figure}

\begin{figure}
 \centering
 \includegraphics[totalheight=5cm,width=6cm]{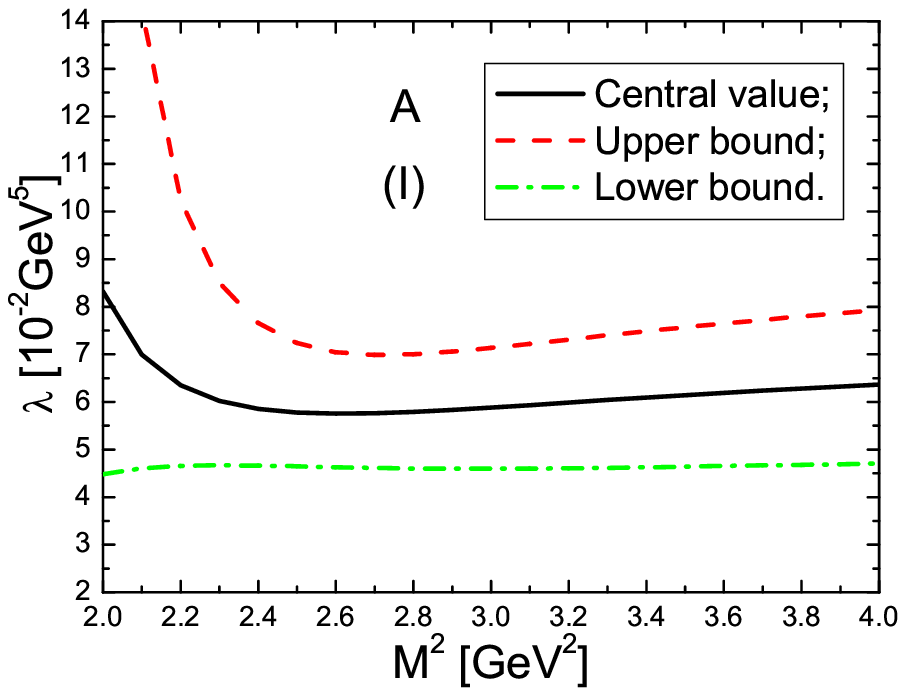}
 \includegraphics[totalheight=5cm,width=6cm]{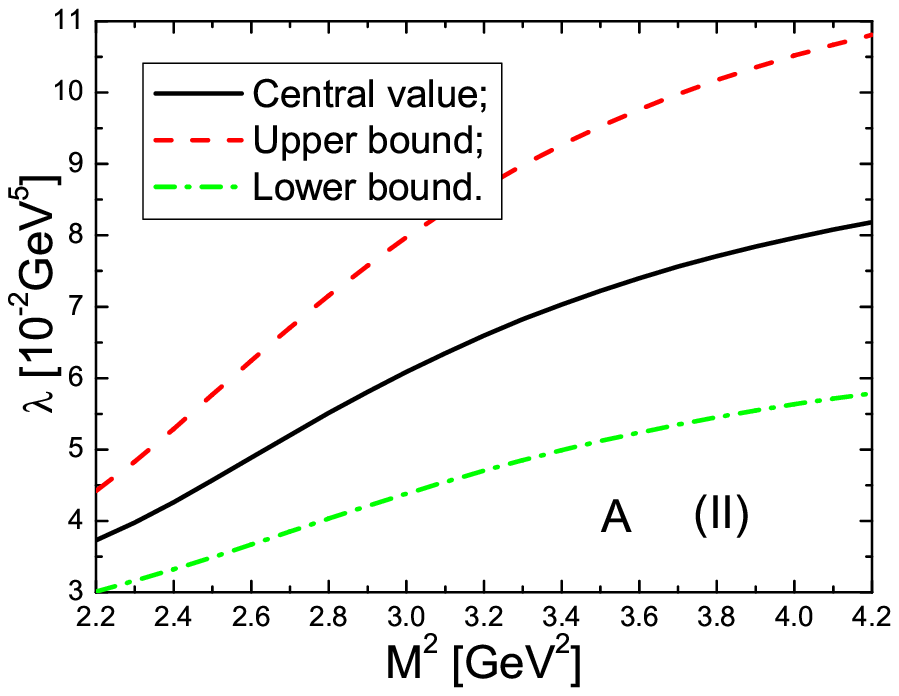}
  \includegraphics[totalheight=5cm,width=6cm]{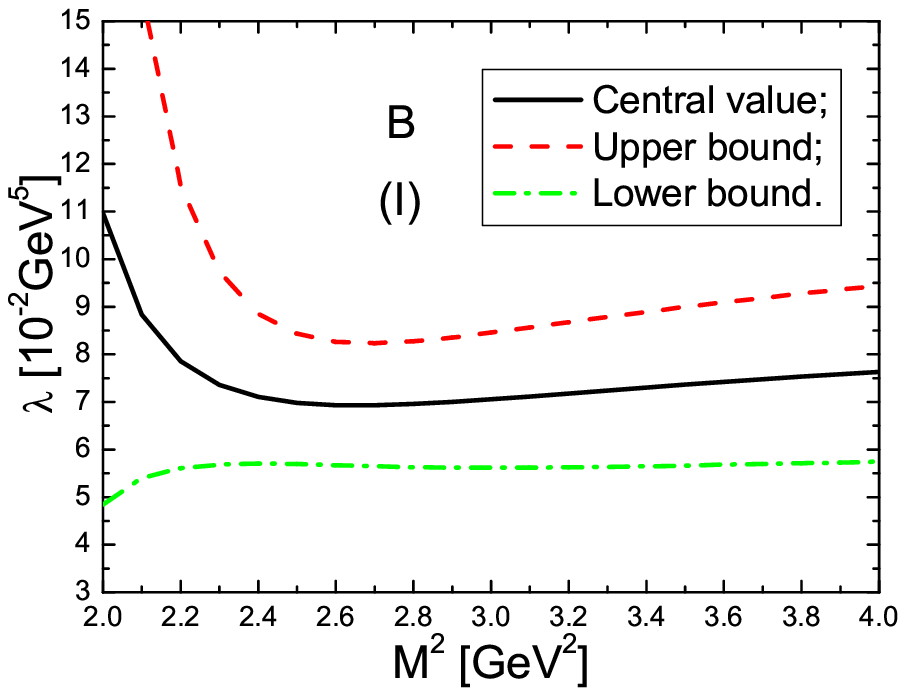}
   \includegraphics[totalheight=5cm,width=6cm]{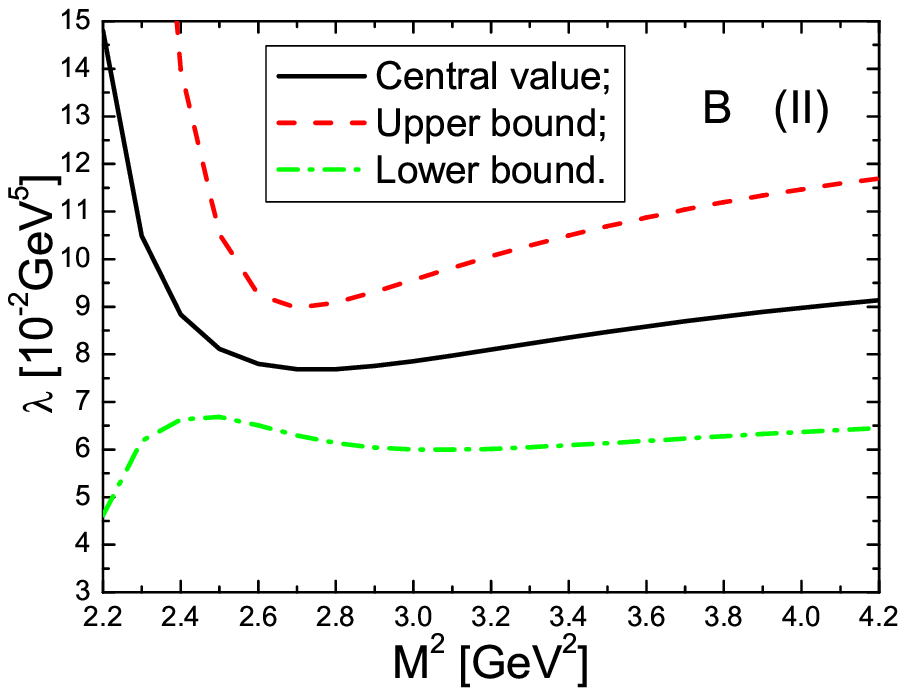}
 \includegraphics[totalheight=5cm,width=6cm]{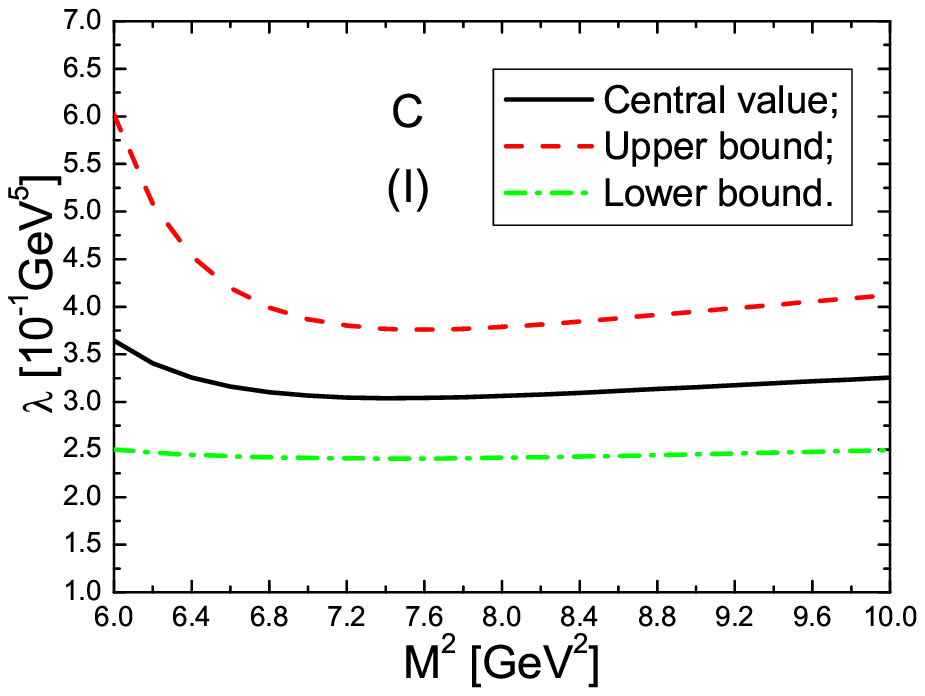}
  \includegraphics[totalheight=5cm,width=6cm]{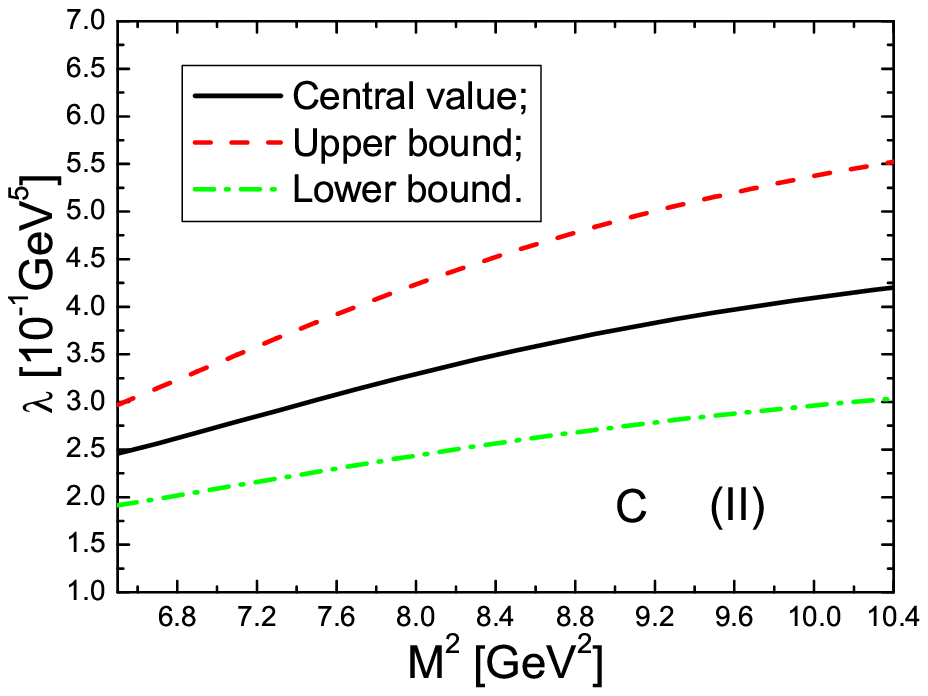}
  \includegraphics[totalheight=5cm,width=6cm]{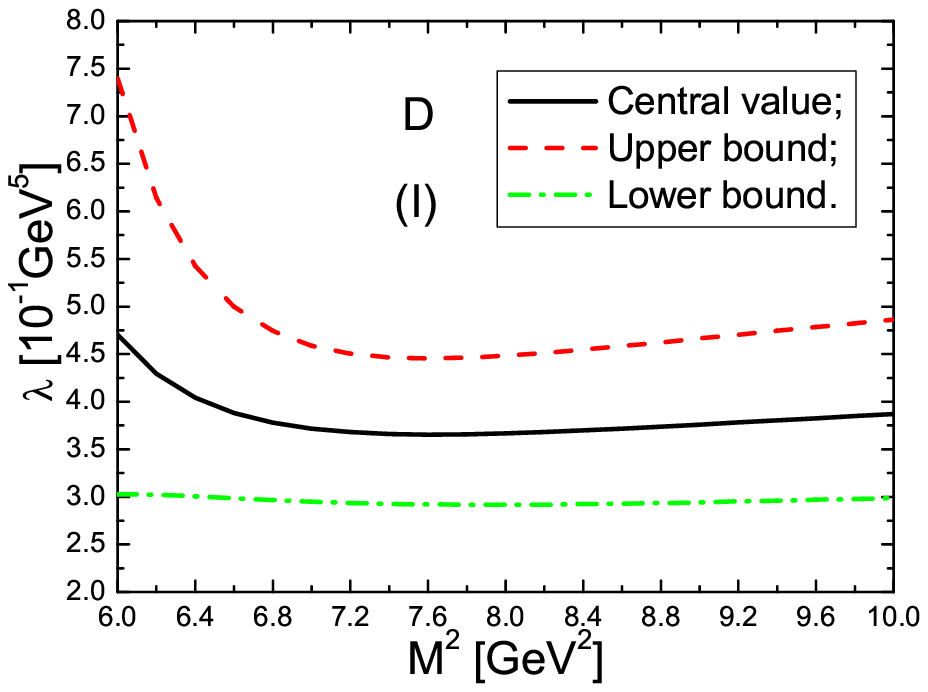}
  \includegraphics[totalheight=5cm,width=6cm]{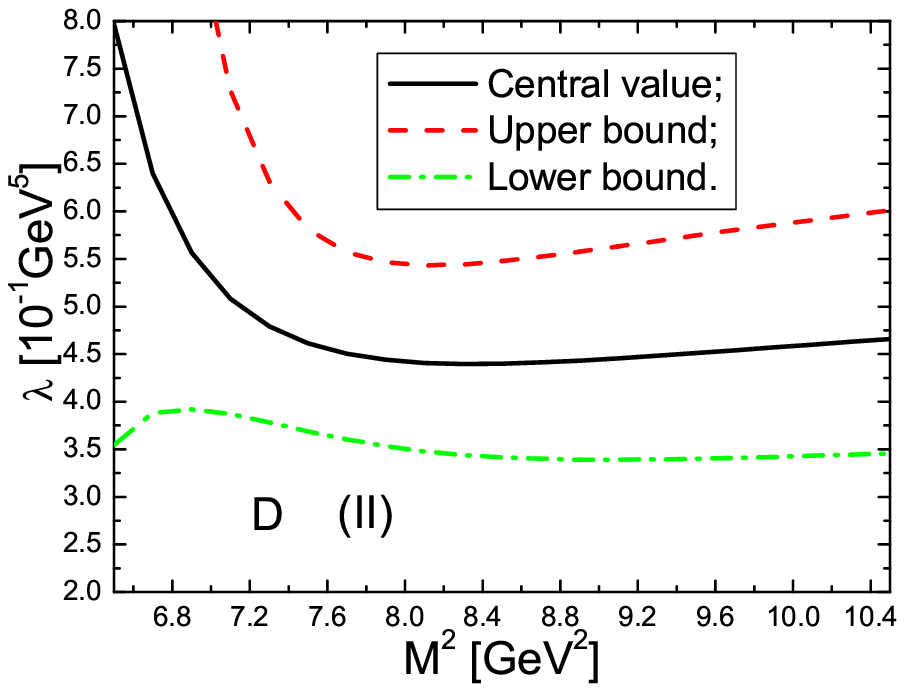}
    \caption{ The pole residues of the scalar tetraquark  states with variation of the Borel parameter $M^2$.
    The $A$, $B$, $C$,
   and $D$ denote the $c\bar{c}q\bar{q}$,
    $c\bar{c}s\bar{s}$, $b\bar{b}q\bar{q}$,
   and $b\bar{b}s\bar{s}$ channels respectively. The (I) and (II) denote the $C\gamma_\mu-C\gamma^\mu$ type and
   the $C\gamma_\mu\gamma_5-C\gamma^\mu\gamma_5$
   type interpolating currents respectively. }
\end{figure}

\section{Conclusion}
In this article, we study the mass spectrum of the scalar hidden
charm and hidden bottom tetraquark states which consist of the
axial-axial type and the vector-vector type diquark pairs with the
QCD sum rules, and observe that the scalar-scalar type and the
axial-axial type tetraquark states  have almost the same ground
state masses while the vector-vector type tetraquark states have
slightly  larger ground state masses. Furthermore, we  compare the
present predictions  with the corresponding ones from a relativistic
quark model based on a quasipotential approach in QCD, and discuss
the values from the constituent diquark model
 based  on  the constituent diquark masses and the spin-spin
 interactions. We can search for
the scalar hidden charm and bottom tetraquark states at the LHCb.

We can perform  Fierz re-ordering in both the Dirac spin space and
the color space to express the tetraquark currents $J(x)$ and
$\eta(x)$ into  a series of $S-S$, $P-P$, $V-V$, $A-A$, $S^i-S^i$,
$P^i-P^i$, $V^i-V^i$ and $A^i-A^i$ color-singlet and color-triplet
meson-meson type currents, there are  contributions from  the
two-particle and many-particle reducible states, those
contaminations  are supposed to be small enough to be neglected
safely. In fact, those contributions maybe considerable (and even
out of control) and impair the predictive  ability. In this article,
we take the single pole approximation for the hadronic spectral
densities, our predictions depend heavily on  the two criteria (pole
dominance and convergence of the operator product expansion) of the
QCD sum rules; the numerical results are rather good.

\section*{Acknowledgements}
This  work is supported by National Natural Science Foundation,
Grant Number 10775051, and Program for New Century Excellent Talents
in University, Grant Number NCET-07-0282.

\section*{Appendix}
The spectral densities at the level of the quark-gluon degrees of
freedom:
\begin{eqnarray}
\rho_{\pm}(s)&=&\frac{1}{256 \pi^6}
\int_{\alpha_{i}}^{\alpha_{f}}d\alpha \int_{\beta_{i}}^{1-\alpha}
d\beta
\alpha\beta(1-\alpha-\beta)^3(s-\widetilde{m}^2_Q)^2(7s^2-6s\widetilde{m}^2_Q+\widetilde{m}^4_Q)
\nonumber \\
&&+\frac{1}{256 \pi^6} \int_{\alpha_{i}}^{\alpha_{f}}d\alpha
\int_{\beta_{i}}^{1-\alpha} d\beta
\alpha\beta(1-\alpha-\beta)^2(s-\widetilde{m}^2_Q)^3(3s-\widetilde{m}^2_Q)
\nonumber \\
&&\pm\frac{ m_sm_Q}{128 \pi^6} \int_{\alpha_{i}}^{\alpha_{f}}d\alpha
\int_{\beta_{i}}^{1-\alpha} d\beta (\alpha+\beta)
(1-\alpha-\beta)^2(s-\widetilde{m}^2_Q)^2(5s-2\widetilde{m}^2_Q)   \nonumber\\
&&+\frac{ m_s\langle \bar{s}s\rangle}{8 \pi^4}
\int_{\alpha_{i}}^{\alpha_{f}}d\alpha \int_{\beta_{i}}^{1-\alpha}
d\beta \alpha \beta
(1-\alpha-\beta)(10s^2-12s\widetilde{m}^2_Q+3\widetilde{m}^4_Q)   \nonumber\\
&&+\frac{ m_s\langle \bar{s}s\rangle}{8 \pi^4}
\int_{\alpha_{i}}^{\alpha_{f}}d\alpha \int_{\beta_{i}}^{1-\alpha}
d\beta \alpha \beta(s-\widetilde{m}^2_Q)(2s-\widetilde{m}^2_Q)   \nonumber\\
&&\mp\frac{ m_Q\langle \bar{s}s\rangle}{8 \pi^4}
\int_{\alpha_{i}}^{\alpha_{f}}d\alpha \int_{\beta_{i}}^{1-\alpha}
d\beta (\alpha+\beta)
(1-\alpha-\beta) (s-\widetilde{m}^2_Q) (2s-\widetilde{m}^2_Q) \nonumber\\
&& \pm\frac{ m_Q\langle \bar{s}g_s\sigma Gs\rangle}{32 \pi^4}
\int_{\alpha_{i}}^{\alpha_{f}}d\alpha \int_{\beta_{i}}^{1-\alpha}
d\beta (\alpha+\beta)
(3s-2\widetilde{m}^2_Q)  \nonumber\\
&&-\frac{ m_s\langle \bar{s}g_s\sigma Gs\rangle}{8 \pi^4}
\int_{\alpha_{i}}^{\alpha_{f}}d\alpha \int_{\beta_{i}}^{1-\alpha}
d\beta \alpha \beta
\left[2s-\widetilde{m}^2_Q+\frac{s^2}{6}\delta(s-\widetilde{m}^2_Q)\right]   \nonumber\\
&& -\frac{ m_sm_Q^2\langle \bar{s}s\rangle}{2 \pi^4}
\int_{\alpha_{i}}^{\alpha_{f}}d\alpha \int_{\beta_{i}}^{1-\alpha}
d\beta(s-\widetilde{m}^2_Q)  \nonumber\\
&&-\frac{ m_s\langle \bar{s}g_s\sigma Gs\rangle}{48 \pi^4}
\int_{\alpha_{i}}^{\alpha_{f}}d\alpha \alpha (1-\alpha)
(3s-2\widetilde{m}^2_Q)   \nonumber\\
&&+\frac{m_Q^2  \langle \bar{s}s\rangle^2}{3 \pi^2}
\int_{\alpha_{i}}^{\alpha_{f}} d\alpha +\frac{m_sm_Q^2\langle
\bar{s}g_s\sigma Gs\rangle }{8 \pi^4}
\int_{\alpha_{i}}^{\alpha_{f}} d\alpha \nonumber\\
&&\mp\frac{m_sm_Q  \langle \bar{s}s\rangle^2}{12 \pi^2}
\int_{\alpha_{i}}^{\alpha_{f}} d\alpha
\left[2+s\delta(s-\widetilde{\widetilde{m}}^2_Q) \right]\nonumber\\
&&-\frac{m_Q^2  \langle \bar{s}s\rangle\langle \bar{s}g_s \sigma
Gs\rangle }{6 \pi^2} \int_{\alpha_{i}}^{\alpha_{f}} d\alpha
\left[1+\frac{s}{M^2} \right]\delta(s-\widetilde{\widetilde{m}}^2_Q)\nonumber\\
&&\pm\frac{5m_s m_Q  \langle \bar{s}s\rangle\langle \bar{s}g_s
\sigma Gs\rangle }{72 \pi^2} \int_{\alpha_{i}}^{\alpha_{f}} d\alpha
\left[1+\frac{s}{M^2} +\frac{s^2}{2M^4}\right]\delta(s-\widetilde{\widetilde{m}}^2_Q)\nonumber\\
&&+\frac{ m_Q^2  \langle \bar{s}g_s \sigma Gs\rangle^2 }{48 \pi^2
M^6} \int_{\alpha_{i}}^{\alpha_{f}} d\alpha
s^2\delta(s-\widetilde{\widetilde{m}}^2_Q)\nonumber\\
 &&\mp\frac{
m_sm_Q \langle \bar{s}g_s \sigma Gs\rangle^2 }{288 \pi^2 M^8}
\int_{\alpha_{i}}^{\alpha_{f}} d\alpha
s^3\delta(s-\widetilde{\widetilde{m}}^2_Q)\, ,
\end{eqnarray}
where $\alpha_{f}=\frac{1+\sqrt{1-4m_Q^2/s}}{2}$,
$\alpha_{i}=\frac{1-\sqrt{1-4m_Q^2/s}}{2}$, $\beta_{i}=\frac{\alpha
m_Q^2}{\alpha s -m_Q^2}$,
$\widetilde{m}_Q^2=\frac{(\alpha+\beta)m_Q^2}{\alpha\beta}$,
$\widetilde{\widetilde{m}}_Q^2=\frac{m_Q^2}{\alpha(1-\alpha)}$, and
$\Delta=4(m_Q+m_s)^2$.

\end{document}